\documentclass[prl,twocolumn,showpacs,superscriptaddress,aps,longbibliography]{revtex4-2}

\usepackage[pdftex]{graphicx}
\usepackage{amsbsy,amssymb,amsmath,bm,mathtools}

\usepackage{xspace}
\usepackage{bm}% bold math
\usepackage{color}

\usepackage{url}
\usepackage[section]{placeins}
\usepackage[colorlinks=true,linkcolor=blue,citecolor=blue]{hyperref}

\usepackage{titlesec}
\titleformat{\section}{\raggedright\bfseries}{\arabic{section}.}{1em}{}
\titleformat{\subsection}{\raggedright\bfseries}{\arabic{subsection}.}{1em}{}

\titlespacing\section{0pt}{12pt plus 4pt minus 2pt}{1pt plus 2pt minus 2pt}
\titlespacing\subsection{0pt}{12pt plus 4pt minus 2pt}{2pt plus 2pt minus 2pt}

\begin{document}

\title{Kibble-Zurek Mechanism for Nonequilibrium Generation \\ of  Magnetic Monopoles in Spin Ices}

\author{Zhijie Fan}
\affiliation{Department of Physics, University of Virginia, Charlottesville, VA 22904, USA}
\affiliation{Department of Modern Physics, University of Science and Technology of China, Hefei, Anhui 230026, China}
\affiliation{Hefei National Laboratory for Physical Sciences at the Microscale, University of Science and Technology of China, Hefei 230026, China}

\author{Adolfo del Campo}
\affiliation{Department  of  Physics  and  Materials  Science,  University  of  Luxembourg,  L-1511  Luxembourg, Luxembourg}
\affiliation{Donostia International Physics Center,  E-20018 San Sebasti\'an, Spain}

\author{Gia-Wei Chern}
\affiliation{Department of Physics, University of Virginia, Charlottesville, VA 22904, USA}

\begin{abstract}
The proliferation of topological defects is a common out-of-equilibrium phenomenon when a system is driven into a phase of broken symmetry. The Kibble-Zurek mechanism (KZM) provides a theoretical framework for the critical dynamics and generation of topological defects in such scenarios. One of the early applications of KZM is the estimation of heavy magnetic monopoles left behind by the cosmological phase transitions in the early universe. The scarcity of such relic monopoles, which contradicts the prediction of KZM, is one of the main motivations for cosmological inflationary theories. On the other hand, magnetic monopoles as emergent quasi-particles have been observed in spin ices, a peculiar class of frustrated magnets that remain disordered at temperatures well below the energy scale of exchange interaction. Here we study the annihilation dynamics of magnetic monopoles when spin ice is cooled to zero temperature in a finite time. Through extensive Glauber dynamics simulations, we find that the density of residual monopole follows a power law dependence on the annealing rate. A kinetic reaction theory that precisely captures the annihilation process from Monte Carlo simulations is developed. We further show that the KZM can be generalized to describe the critical dynamics of spin ice, where the exponent of the power-law behavior is determined by the dynamic critical exponent $z$ and the cooling protocol. 
\end{abstract}

\maketitle

The existence of a critical point has profound implications on the properties of a system, both in and out of equilibrium. In particular, crossing a continuous phase transition in a finite time leads to breaking adiabatic dynamics. As a result, topological defects proliferate in the driven system. In this context, the Kibble-Zurek mechanism (KZM) provides a reference theoretical framework for critical dynamics~\cite{Kibble76a,Kibble76b,Zurek96a,Zurek96c,DZ14}. It unveils that the latter behavior is universal and characterized by scaling laws that govern the density of defects and the response time of the driven system. In particular, KZM has been employed to understand the formation of 't Hooft-Polyakov magnetic monopoles, a topological defect of non-abelian gauge theories, in the early universe~\cite{Zeldovich78,Preskill79}. The experimental absence of such fundamental magnetic monopoles  led to the ideas of cosmological inflation~\cite{Einhorn80,Guth81}. On the other hand, condensed matter systems support various emergent topological defects and offer a fruitful arena for examining various aspects of KZM.

Universality away from equilibrium can be brought out by considering a system in which different phases of matter are accessible by varying an external control parameter $\lambda$ (temperature, density, etc.) across a critical value~$\lambda_c$. A continuous phase transition is characterized by a universal equilibrium scaling law of the correlation length $\xi=\xi_0/|\epsilon|^\nu$, where $\epsilon=(\lambda-\lambda_c)/\lambda_c$ and $\nu$ is the correlation-length critical exponent. Similarly, the equilibrium relaxation time diverges in the neighborhood of the critical point~$\lambda_c$ as $\tau=\tau_0/|\epsilon|^{z\nu} \sim \xi^z$, where $z$ is the dynamic critical exponent. This divergence is known as critical slowing down and is responsible for breaking adiabaticity in any finite-time driven protocol $\lambda(t)$. To appreciate this, it suffices to linearize $\lambda(t)$ in the neighborhood of $\lambda_c$ so that $\epsilon=t/\tau_Q$, assuming that the critical point is reached at $t=0$. The KZM predicts that the density of point-like defects in $d$ spatial dimensions scales as $n\sim\hat{\xi}^{-D}$, where $D$ is the spatial dimension, and $\hat{\xi}$ is the non-equilibrium correlation length $\hat{\xi}=\xi_0(\tau_Q/\tau_0)^{\frac{\nu}{1+z\nu}}$ which exhibits a power-law scaling with the quench time $\tau_Q$ that is fixed by the equilibrium critical exponents $z$ and~$\nu$. An additional prediction of the KZM is that the characteristic response time, known as the freeze-out-time $\hat{t}$, also scales universally with the quench time $\tau_Q$ as $\hat{t}=(\tau_0\tau_Q^{z\nu})^{\frac{1}{1+z\nu}}$. These predictions can alternatively be derived using finite-time scaling \cite{Chandran12,Nikoghosyan16}.

The nonequilibrium critical behavior predicted by the KZM has been explored in depth in one-dimensional systems \cite{Laguna98,Antunes06,Zurek09,Suzuki09a,delcampo10,Das12,Sonner15,GRMdC19}. The spatial distribution of topological defects is then highly constrained, and exact analytical descriptions are often possible. Experimental evidence is convincing in the quantum domain \cite{Guo14,Wang14,Wu16,Cui16,Lukin17,Bando20,King2022} but remains limited in systems admitting a classical description  \cite{Monaco02,EH13,Ulm13,Pyka13,Lamporesi13}.
Results in higher spatial dimensions show a rich behavior. In theoretical and experimental studies,  some settings are consistent with the scaling predictions dictated by the KZM \cite{Casado01,Casado06,Chae12,Lin14,Navon15,Chomaz15,Meier17,Shin19,Goo21,KimShin22,Du23},  while others display deviations \cite{Jeli11,Griffin12,Keim15,Chesler15,Du23}. The critical dynamics in systems with a complex vacuum manifold supporting different kinds of topological defects remains poorly understood, as coarsening and multiple channels for defect creation and annihilation can coexist~\cite{Biroli10,Libal20}.

\begin{figure*}
    \includegraphics[width=1.8\columnwidth]{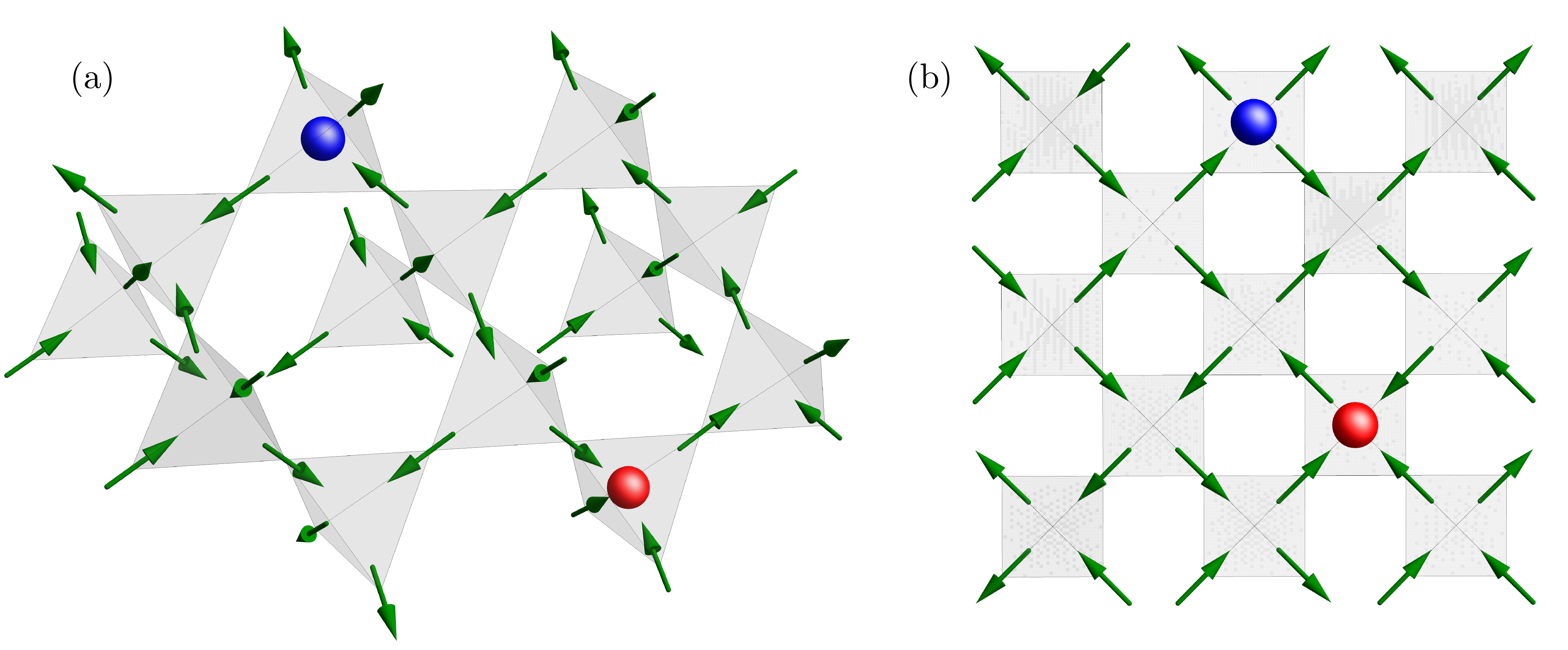}
    \caption{Schematic representation of (a) the pyrochlore spin ice and (b) 2D spin ice on the checkerboard lattice. The red (blue) sphere represents topological defects, which are magnetic monopoles in spin ices and kinks/anti-kinks in Ising chain.}
    \label{fig:spin-ice}
\end{figure*}

Spin-ice systems~\cite{Snyder04,Harris97,Bramwell01,Hertog00} are an unusual class of ferromagnet where the magnetic atoms reside on a pyrochlore lattice, a three-dimensional network of corner-sharing tetrahedra as shown in FIG.~\ref{fig:spin-ice}(a). For spin ice with interactions restricted to nearest neighbors, the magnet remains in a disordered state down to zero temperature. At first sight, the KZM is not expected to describe the annealing dynamics of such idealized spin ice, which shows no symmetry breaking. However, at temperatures below the energy scale of exchange interaction, spin ice exhibits novel fractionalized quasi-particles which carry a net magnetic charge, essentially behaving as magnetic monopoles~\cite{Castelnovo2008}. Conservation of  magnetic charges means that these quasi-particles have to be created and annihilated in pairs. Magnetic monopoles are thus topological defects in an otherwise disordered spin state, in contrast to topological defects due to broken symmetry as in standard KZ scenario.  An intriguing question is whether these emergent magnetic monopoles in a quenched spin ice exhibit scaling behaviors and if the KZM can be generalized to describe their nonequilibrium dynamics.

The emergence of magnetic monopoles in spin ice is closely related to the ice rule, a local constraint for ground states. Dominant easy-axis anisotropy forces the magnetic moments to point in the local $\langle 111 \rangle$ directions, allowing us to express spins in terms of Ising variables: $\mathbf S_i = \sigma_i \mu \hat{\mathbf e}_i$, where $\mu$ is the magnitude of the magnetic moment, $\hat{\mathbf e}_i$ is the local crystal-field axis, and $\sigma_i = \pm 1$ indicates the direction of the magnetic moment, which points either from the center of a tetrahedron to the corresponding corner or vice versa. Both the short-range ferromagnetic exchange $J_F < 0$ and the long-range dipolar interaction contribute to an effective nearest-neighbor antiferromagnetic interaction between the Ising spins $\mathcal{H} = J \sum_{\langle ij \rangle} \sigma_i \sigma_j$, where $J = \frac{1}{3} (|J_F| \mu^2 + 5\mu_0 \mu^2 /4\pi a^3) $ is the effective antiferromagnetic interaction and $a$ is the nearest-neighbor distance in pyrochlore lattice. We first focus on the annealing dynamics with interactions restricted to nearest neighbors and discuss effects of long-range dipolar interaction later.

It is convenient to express the spin-ice energy in terms of magnetic charges for understanding the ground-state properties and elementary excitations. To this end, we use the dumbbell approximation~\cite{Castelnovo2008} to replace a magnetic moment $\mathbf S_i$ (a dipole) by two opposite magnetic charges $\pm \mu/ \ell$ at the two ends of a bar of length $\ell$, which is set to be the distance between centers of two nearest-neighbor tetrahedra. The effective magnetic charge of a tetrahedron-$\alpha$ is then $Q_{\alpha} = \pm (\mu/\ell) \sum_{i \in \alpha} \sigma_i$, where the $\pm$ sign is used for tetrahedra of opposite orientations, and the sum is over the four spins of the tetrahedron. In terms of magnetic charges, the system energy becomes $\mathcal{H} = \frac{v}{2} \sum_{\alpha} Q_{\alpha}^2$ up to an irrelevant constant, where the self-energy coefficient $v = J \ell^2/\mu^2$.
The total energy of a spin ice is thus minimized by any spin configurations with zero magnetic charges $Q_{\alpha} = 0$ for all tetrahedra, which form a diamond lattice that is dual to the pyrochlore lattice. The charge neutral condition corresponds to a tetrahedron with two $\sigma=+1$ and two $\sigma=-1$ Ising spins, known as the 2-in-2-out ice rules~\cite{Bramwell01}. While these constraints introduce strong short-range correlations between spins, no long-range order is induced even at zero temperature. The number of ground states satisfying the ice rules grows exponentially with the system size, giving rise to a zero-point entropy, which is well approximated by the Pauling estimate $S_{\rm Pauling} = (1/2) \log(3/2)$ and verified experimentally in canonical spin ice compounds~\cite{Pauling35,Ramirez99}.

Elementary excitations above the hugely degenerate ground-state manifold are represented by tetrahedra that violate the ice rules~\cite{Castelnovo2008,Jaubert_2009}. These correspond to 3-in-1-out/1-in-3-out tetrahedra with a magnetic charge $Q = \pm q_m$, or 4-in/4-out tetrahedra with charge $Q = \pm 2 q_m$, where $q_m = 2 \mu/\ell$ is the elementary unit of magnetic charges in spin ice. These defect tetrahedra, particle-like objects carrying net magnetic charges, are essentially magnetic monopoles. It is also worth noting that the monopoles in spin ice are topological defects as they have to be created and annihilated in pairs. For example, a single spin-flip, or an inverted dumbbell, results in two monopoles of charge $Q = \pm q_m$ on adjacent diamond-lattice sites. Crucially, the monopoles can be separated from one another without further violations of local neutrality by flipping a chain of adjacent dumbbells.

The vacuum of these emergent magnetic monopoles corresponds to the highly constrained ground-state manifold. It has been shown that an effective magnetostatic theory can describe this manifold. Indeed, monopole excitations are the source and sink of the emergent magnetic field $\mathbf B(\mathbf r)$. The ice rules, i.e., the absence of monopoles, translate to the divergence-free condition $\nabla \cdot \mathbf B = 0$, which in turn gives rise to dipolar-like power-law spin correlations in the degenerate ground-state manifold~\cite{isakov04,henley05,Ryzhkin05}. The monopole density determines the correlation length $\xi$ of this emergent critical state at $T \to 0$: $\xi \sim 1/n_m^{1/3}$. As an activation energy $\Delta E_{m} = \frac{v}{2} q_m^2 = 2J$ is required to create fundamental monopoles of charge $\pm q_m$, the density of such topological defects $n_m \sim e^{-2J/T}$ is exponentially suppressed at low temperatures. This results in an equilibrium correlation length $\xi \sim e^{2J/3T}$, which diverges exponentially as $T \to 0$, in contrast to the familiar power-law divergence when approaching a conventional critical point.

\begin{figure*}[ht]
\includegraphics[width=0.475\linewidth]{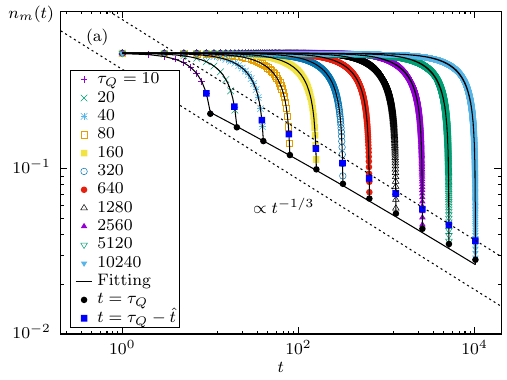}
\includegraphics[width=0.475\linewidth]{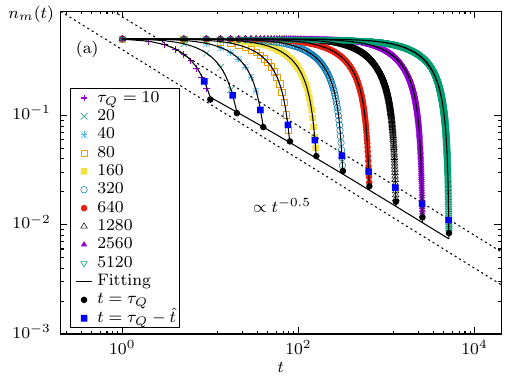}
\caption{Density of magnetic monopoles $n_m$ versus time during (a) linear cooling $\gamma(t) = t/\tau_Q$ and (b) algebraic cooling with $\alpha = 2$ for pyrochlore spin ice. The solid black curves denote results obtained from the numerical integration of the rate equations. The densities at the end of cooling $t = \tau_Q$ are shown as solid black circles, while those at the freeze-out time $t = \tau_Q - \hat{t}$ are denoted by solid black squares. The monopole densities at both $\tau_Q$ and the freeze-out time exhibit power-law behaviors with the same exponent in the large $\tau_Q$ limit. The dashed lines are the KZM prediction Eq.~(\ref{eq:KZM-nm}).}
    \label{fig:nm-vs-t}
\end{figure*}

A similar exponential divergence of correlation length also occurs in the paradigmatic ferromagnetic 1D Ising model. Similar to spin ices, Ising spins remain disordered at any finite temperature.  An unconventional critical point at $T_c = 0$ can be associated with the system, at which spins become fully polarized. The average distance between kink and anti-kink pairs, which are topological defects of an Ising chain, determines the correlation length. The fact that the number of kinks is suppressed at low-$T$ similarly gives rise to a correlation length that grows exponentially as $T \to 0$. 
Moreover, the dynamical behavior of the 1D Ising model under the Glauber dynamics can be described by a solvable master equation~\cite{Glauber1963, NaimBook, Lee2020, Krapivsky2010}. Notably, the KZ scaling hypothesis has also been verified in the 1D Ising model when the system is slowly annealed to zero temperature~\cite{Adolfo2021,Lee2020}.

From the viewpoint of an unconventional critical point at $T_c = 0$, spin ices can be viewed as a different high-dimensional generalization of the 1D Ising chain, to be contrasted with the standard square or cubic Ising models. We note that a 2D analog of the pyrochlore spin ice is given by the antiferromagnetic Ising model on a checkerboard lattice, as shown in FIG.~\ref{fig:spin-ice}(b). An artificial version of such 2D spin ice has been realized in arrays of nanomagnets~\cite{Wang06,Nisoli13,Chern14,Perrin16} and optical traps of soft-matter particles~\cite{Libal06,Ortiz16,Ortiz19}. Despite the similarity, we note that while the Ising chain becomes long-range ordered at $T=0$, both spin ices remain disordered down to zero temperature when interactions are restricted to nearest neighbors. Here we show that KZM, with proper modification, can also be applied to the critical dynamics of spin ice and the annihilation of magnetic monopoles.

\subsection{Annealing of spin ice with Glauber dynamics}

\noindent 
To describe the nonequilibrium dynamics associated with a temperature quench, we perform Glauber dynamics ~\cite{Glauber1963, NaimBook} simulations of pyrochlore spin ice with time-dependent temperature $T(t)$. To take into account the stochastic and local nature of the spin dynamics, at each fundamental step, a spin $\sigma_i$ that is randomly chosen from the system is updated according to the transition probability $w(\sigma_i \to -\sigma_i) = \frac{1}{2} [1 - \textrm{tanh}\left(\frac{1}{2}\beta \Delta E_i\right) ]$, where $\beta = 1/T$ is inverse temperature and $\Delta E_i$ is the energy change due to the flipped spin. At low temperatures, a single-spin flip results in mostly either the creation/annihilation of monopole pairs, of which $\Delta E= \pm 4J$, or the movement of monopoles for which $\Delta E = 0$. It is thus convenient to introduce a dimensionless parameter $\gamma(t) = \tanh[2\beta(t) J]$ which controls the transition rate. For example, ignoring the updates that involve double monopoles, the transition rate at low temperatures simplifies to $w(\sigma_i \to -\sigma_i; t) = \frac{1}{2} [1-\gamma(t) \sigma_i \, {\rm sign}(h_i)]$, where $h_i = \sum_{j \in {\rm nn}(i)} \sigma_j$ is the sum of nearest-neighbor Ising spins, and ${\rm sign}(x)$ is the sign function.

In terms of this control parameter, we first consider the so-called linear cooling schedule: $\gamma(t) = t / \tau_Q$, where $\tau_Q$ denotes the total annealing time~\cite{Krapivsky2010}. With this cooling protocol, the system evolves from $T = \infty$ at $t=0$ to zero temperature when $t = \tau_Q$. 
The time is incremented by $\delta t = 1/N_s$ after each spin update attempt, where $N_s = 16 L^3$ is the total number of spins in the system. All simulations below were performed on a lattice of $L = 10$, with $N_s =16,000$ spins. After one Monte Carlo sweep of the entire system, the time increases by one unit of time $\Delta t = 1$, and the charge statistics is measured. The cooling time varies in the range $\tau_Q = 10\times 2^n$ with $n=0, 1, 2, 3, \dots, 10$. The final results are obtained by averaging the data from $10,000$ randomly generated initial states.

\begin{figure*}[t]
\includegraphics[width=0.475\linewidth]{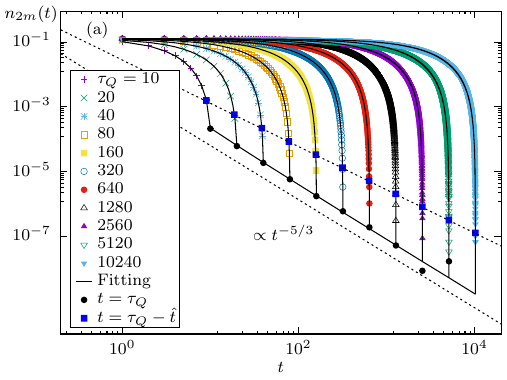}
\includegraphics[width=0.475\linewidth]{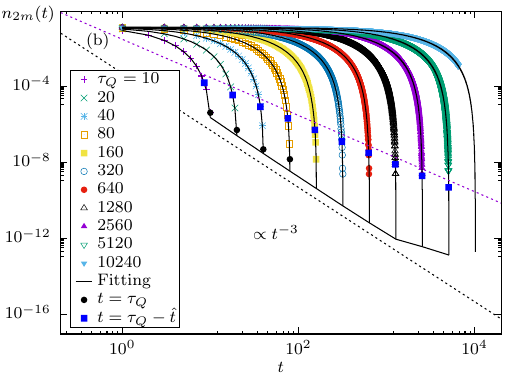}
\caption{Density of double monopoles $n_{2m}$ versus time during (a) linear cooling $\gamma(t) = t/\tau_Q$ and (b) algebraic cooling with $\alpha = 2$ for pyrochlore spin ice. The solid black curves denote results obtained from the numerical integration of the rate equations. The densities at the end of cooling $t = \tau_Q$ are shown as solid black circles, while those at the freeze-out time $t = \tau_Q - \hat{t}$ are denoted by solid black squares. The monopole densities at both $\tau_Q$ and the freeze-out time exhibit a power-law behavior in the large $\tau_Q$ limit. The dashed lines power laws from Eq.~(\ref{eq:n2m-freeze}) and~(\ref{eq:n2m-left}).}
\label{fig:n2m-vs-t}
\end{figure*}

The time dependence of the elementary-monopole density $n_m(t)$ is shown in FIG.~\ref{fig:nm-vs-t} for algebraic cooling with $\alpha = 1$ and 2, and varying cooling time~$\tau_Q$. As discussed above, these are 3-in/1-out or 1-in/3-out tetrahedra carrying a net charge $Q=\pm q_m$, Another type of defect tetrahedra with all spins pointing in or out can be viewed as a quasi-bound state of two fundamental monopoles of equal charges. The density $n_{2m}(t)$ of such double monopoles as a function of time is shown in FIG.~\ref{fig:n2m-vs-t}. As these quasi-bound states carry a doubled charge $Q = \pm 2 q_m$, they are energetically more costly, giving rise to a density that is orders of magnitude smaller than that of monopoles. The critical dynamics of both types of quasi-particles exhibit a similar overall pattern: an initial slow decay that lasts a long time, followed by a very steep decline at the end of cooling.

To shed  light on the annealing dynamics of spin ices, rate equations based on reaction kinetics theory are developed to describe the dynamical evolution of single and double monopoles. For example, the rate equation for magnetic monopoles of charges $\pm q_m$ at the late stage of the cooling is
\begin{eqnarray}
	\label{eq:rate-nm}
	\frac{d n_m}{dt} = \mathcal{A}_0 + \mathcal{A}_1 n_{2m} + \mathcal{A}_2 n_{2m}^2 - \mathcal{B} n_m^2,
\end{eqnarray}
The first three $\mathcal{A}$ terms denote the various mechanisms for producing $\pm q_m$ monopoles: pair-creation from vacuum, decay of a double monopole, and conversion of two double monopoles into fundamental monopoles. The last term accounts for the pair annihilation of $\pm q_m$ monopoles. It is worth noting that the leading decay term is quadratic in $n_m$ (no linear term) is a manifestation of their topological nature. 

Through reaction kinetic theory, the coefficient $\mathcal{B}$ is uniquely related to the three $\mathcal{A}$ coefficients, which will be treated as fitting parameters. In practice, these parameters are determined from Glauber dynamics simulations with a small $\tau_Q =160$. The rate equation for the higher-energy double monopoles $n_{2m}$ can be similarly obtained; see Appendix~B for details. Using random spins to set initial conditions, the rate equations are integrated numerically. The results are shown in FIG.~\ref{fig:nm-vs-t} as solid lines. Remarkably, the reaction kinetics based on exactly the same set of parameters gives an excellent overall agreement with the Glauber dynamics simulations for both linear and algebraic $\alpha = 2$ cooling schedules. 

\subsection{Kibble-Zurek mechanism for monopoles}

\noindent
Both Monte Carlo simulations and calculations using the rate equations yield a power-law dependence on $\tau_Q$ for the residual monopole density at the end of cooling $n_m(\tau_Q) \sim \tau_Q^{-\mu}$, where the KZ exponent is $\mu \approx 0.33$ and 0.5 for algebraic cooling with $\alpha = 1$ and 2, respectively. 
Here we show that these scaling behaviors can be explained by a generalized KZM similar to that adopted for the 1D Ising chain~\cite{Krapivsky2010}. 
As discussed above, although spin-ice systems exhibit a critical point at $T_c =0$,  the correlation length diverges exponentially $\xi \sim n_{m}^{-1/D} \sim e^{\Delta E_{m}/DT}$, instead of algebraically  as in a conventional critical point. Here the spatial dimension $D = 2$ and 3 for the checkerboard and pyrochlore spin ice, respectively, and the activation energy $\Delta E_{m} = 2J$ for both cases. On the other hand,  the relaxation time $\tau$, which is closely related to the annihilation rate of monopoles, also diverges exponentially as $T \to 0$~\cite{Jaubert_2009}. Consequently, one can still define a dynamical exponent that relates these two exponentially divergent quantities:~$\tau \sim \xi^z$. 

The relaxation time is shown to follow the Arrhenius law: $\tau \sim e^{\Delta E_{m}/T} \sim e^{2J/T}$ for spin ice with nearest-neighbor interaction~\cite{Jaubert_2009}. The exponential divergence of the relaxation time has also been explicitly verified from the decay of monopoles in instant-quench simulations; see Appendix~A for details. This gives rise to a dynamical exponent $z = D$ for the relaxation of spin ice, which is also explicitly confirmed in our quench simulations. Finally, we note in passing that, despite the similarity between the Ising chain and spin ices, the dynamical exponent for kinks is $z = 2$ in the 1D Ising model~\cite{Krapivsky2010,Lee2020}.

Central to the KZM is the freeze-out time $\hat{t}$, measured from the critical point, which signifies the breaking of adiabaticity. Before the freeze-out time during the cooling, the system can reach the quasi-equilibrium state of the instantaneous temperature $T(t)$ due to a short relaxation time $\tau(T)$ at the corresponding temperature. Freezing of the system occurs when the exponentially increasing relaxation time is comparable to the time left before reaching the critical point at $T_c = 0$, i.e.
\begin{eqnarray}
	\label{eq:freeze-t}
	\hat{\tau} = \tau\left( T(\tau_Q - \hat{t}) \right) = \hat{t}.
\end{eqnarray}
For time $t \gtrsim \tau_Q - \hat{t}$, breaking adiabaticity means that the pair-annihilation of topological defects is suppressed. The number of monopoles at the end of annealing can thus be well approximated by that at the freeze-out time $n_m(\tau_Q) \sim n_m(\tau_Q - \hat{t})$.
Here we demonstrate the determination of $\hat{t}$ for the general algebraic cooling schedule  
\begin{eqnarray}
	\label{eq:alg-cooling}
	1-\gamma(t) = A \left(1 - \frac{t}{\tau_Q} \right)^\alpha,  
\end{eqnarray}
when $t \to \tau_Q$. Here $A >0$ is a positive constant. The linear cooling schedule corresponds to $\alpha = 1$. Substituting the resultant time-dependent temperature $T(t)$ into Eq.~(\ref{eq:freeze-t}), we have $\hat{t} = \exp\{\tanh^{-1}[1-A(\hat{t}/\tau_Q)^\alpha]\}$. Assuming slow cooling such that $\tau_Q \gg \hat{t}$, we expand the right-hand side of this equation to leading order in $\hat{t}/\tau_Q$, and obtain a scaling relation
\begin{eqnarray}
	\label{eq:hat-t}
	\hat{t} \sim \tau_Q^{\alpha / (2 + \alpha)}.
\end{eqnarray}
The residual density of monopoles can then be estimated from the correlation length at the freeze-out time, i.e., $n_m(\tau_Q)  \sim \hat{\xi}^{-D} \sim \hat{\tau}^{-D/z}$. Remarkably, the fact that the dynamical exponent is given by the dimension of spin ice $z = D$ means that $n_m(\tau_Q) \sim \hat{\tau}^{-1}$, independent of the dimension. Combining the KZ condition~(\ref{eq:freeze-t}) and the scaling of freeze-out time in Eq.~(\ref{eq:hat-t}), we obtain a power-law dependence
\begin{eqnarray}
	\label{eq:KZM-nm}
	n_m(\tau_Q) \sim \tau_Q^{-\alpha / (2 + \alpha)},
\end{eqnarray}
which is independent of spatial dimensions. For $\alpha = 1$ and 2, the above formula gives a KZ exponent $\mu = 1/3$ and 1/2, consistent with our numerical results shown in FIG.~\ref{fig:nm-vs-t}.  Notably, the monopole densities computed at the freeze-out time and at the end of the cooling exhibit the same power-law behavior. Moreover, we have explicitly verified numerically that the same exponents also apply to the 2D checkerboard spin ice subject to algebraic cooling schedules.

\begin{figure*}[t]
    \centering
    \includegraphics[width=0.475\linewidth]{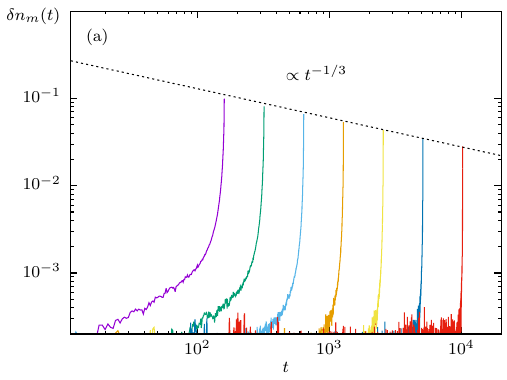}
    \includegraphics[width=0.475\linewidth]{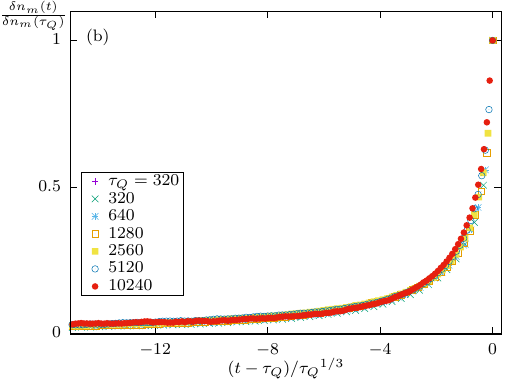}
    \caption{(a) Growth of the excess monopole density for different values of the quench time $\tau_Q$ obtained from Glauber dynamics simulations. (b) The normalized excess monopole density as a function of the evolution time rescaled by the freeze-out $\hat{t}$ shows the dynamic scaling of the universal nonequilibrium behavior. }
    \label{fig:dyn-scaling}
\end{figure*}

\subsection{Residual density of double monopoles}

\noindent    
The double-monopole density $n_{2m}(t)$ obtained from Glauber dynamics simulations is shown in FIG.~\ref{fig:n2m-vs-t} as a function of time. Again, the simulation results are well captured by the rate equations. While the double monopoles seem to also exhibit power-law behavior both at the freeze-out time and at the end of cooling, the two exponents are different contrary to the case of single monopoles. It is worth noting that the double-monopoles are not topological defects as they can spontaneously decay into two fundamental monopoles. As a result, there is no freezing for the annihilation of double monopoles. However, we can still estimate the density of double monopoles at the freeze-out time $t = \tau_Q - \hat{t}$. As the activation energy of such defects is $\Delta E_{2m} = \frac{v}{2} (2 q_m)^2 = 8J$, their equilibrium density scales as $n_{2m} \sim e^{- 8 J/T}$. Since the relaxation time $\tau \sim e^{2 J/T}$ in the adiabatic regime, we have $n_{2m} \sim \tau^{-4}$. Using the KZ condition~(\ref{eq:freeze-t}) that the relaxation time at the freeze-out instant is $\hat{\tau} = \hat{t}$, the density of double monopole at the freeze-out time is 
\begin{eqnarray}
	\label{eq:n2m-freeze}
	n_{2m}(\tau_Q - \hat{t}) \sim  \tau_Q^{-4\alpha/(2+\alpha)}.
\end{eqnarray}
This power law agrees very well with the numerical results for both linear cooling and algebraic cooling with $\alpha = 2$; see FIG.~\ref{fig:n2m-vs-t}.

However, as discussed above, since double monopoles are non-topological, they will continue to decay even after the freeze-out instant. Their relaxation in this regime is governed by a rate equation 
\begin{eqnarray}
	\label{eq:rate-n2m}
	\frac{dn_{2m}}{dt} = \frac{3 n_m^2}{16 \tau_{2m}} e^{-4\beta(t) J}  - \frac{n_{2m} }{ \tau_{2m}},
\end{eqnarray} 
where the temperature-independent $\tau_{2m}$ is the intrinsic lifetime of the double monopole. The first term above describes the combination reaction of two same charge monopoles into a double monopole. The reverse process, corresponding to the second term above, is the dominant contribution to the decay of double monopoles. In this freeze-out regime, the density of fundamental monopoles can be approximated by its value at the freeze-out instant. The depletion of $n_m$ due to the recombination is negligible due to the small exponential factor $e^{-4\beta J}$ at very low temperatures. Assuming a short decay time of double monopoles $\tau_{2m} \ll \hat{t}$, the rate equation for the case of algebraic cooling can be integrated to give a residual density  
\begin{eqnarray}
	\label{eq:n2m-left}
	n_{2m}(\tau_Q) \sim   \tau_Q^{-(4\alpha + \alpha^2)/(2 + \alpha)}.
\end{eqnarray}
Details of the derivation is presented in Appendix~C. This power law dependence is confirmed by both Glauber dynamics simulations and rate equation, as shown in FIG.~\ref{fig:n2m-vs-t}.

\subsection{Dynamical scaling}

\noindent
The freeze-out time $\hat{t}$ and the associated correlation length of KZM also provide a basis for dynamically scaling the nonequilibrium behavior during cooling~\cite{Francuz16,Lee2019,Lee2020,Chandran12,Nikoghosyan16}. In particular, here we consider the time-dependent excess monopole density defined as $\delta n_m(t) = n_m(t) - n_m^{\rm (eq)}(t)$, which represents a genuine nonequilibrium part of the defect density. Here the quasi-equilibrium monopole density is given by the Boltzmann distribution at the instantaneous temperature $n_m^{\rm (eq)}(t) \sim \exp[-\beta(t) \Delta E_{m}]$ with the degeneracy factor adequately taken into account. The excess monopole density as a function of time is shown in the inset of FIG.~\ref{fig:dyn-scaling}(a) for various cooling rates. The density of excess monopoles becomes non-zero immediately after the cooling, yet remains rather small, of the order of $\delta n_m \sim 10^{-3}$, in the initial quasi-adiabatic regime. During this period, the number of excess monopoles increases gradually until the freeze-out time, which is marked by the abrupt, rapid growth of $\delta n_m$. At the end of cooling, when the system reaches zero temperature and $n^{\rm (eq)}_m = 0$, the density of excess defects exhibits the same scaling $\delta n_m \sim \tau_Q^{-1/3}$ as shown by the dashed line. 

\begin{figure*}[ht]
\includegraphics[width=0.475\linewidth]{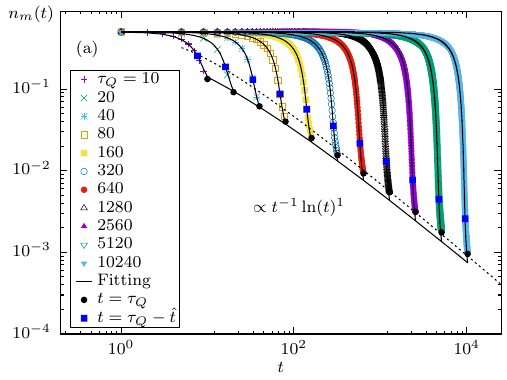}
\includegraphics[width=0.475\linewidth]{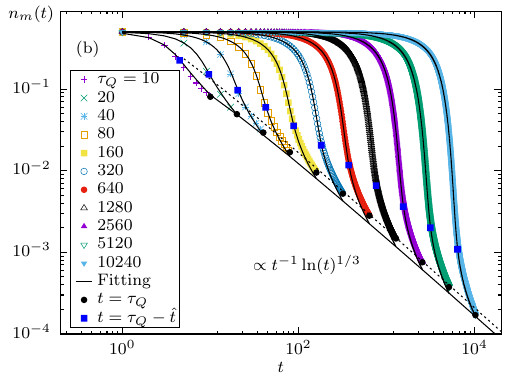}
\caption{Density of magnetic monopoles $n_m$ versus time during exponential cooling with (a) $\alpha = 1$ and (b) $\alpha = 3$ for pyrochlore spin ice. The solid black curves denote results obtained from the numerical integration of the rate equations. The densities at the end of cooling $t = \tau_Q$ are shown as solid black circles, while those at the freeze-out time $t = \tau_Q - \hat{t}$ are denoted by solid black squares. The monopole densities at both $\tau_Q$ and the freeze-out time exhibit power-law behaviors with the same exponent in the large $\tau_Q$ limit. The dashed lines are the KZM prediction Eq.~(\ref{eq:nm-left-exp}).}
    \label{fig:nm-vs-t-exp}
\end{figure*}

It is worth noting that the relevant time scale that determines the evolution of the quenched system is the freeze-out time instead of the annealing time $\tau_Q$. The dynamical scaling posits that the density of excess monopoles, normalized by the density of residual defects at the end of cooling, is a universal function of the time left before reaching the critical point, rescaled by the freeze-out time
\begin{eqnarray}
	\delta n_m(t) = \delta n_m(\tau_Q) \mathcal{F}\!\left(\frac{t-\tau_Q}{\hat{t}} \right).
\end{eqnarray}
The critical point $t = \tau_Q$ corresponds to $\mathcal{F}(0) = 1$. As shown in FIG.~\ref{fig:dyn-scaling}(b), the rescaled data points from our Glauber dynamics simulations collapse on a universal curve, underscoring a universal nonequilibrium dynamical behavior.

\subsection{Exponential cooling protocol}
\label{sec:exponential-cooling}

\noindent Since either the Glauber or Metropolis dynamics for Ising spins is controlled by the Arrhenius factor $e^{-4\beta J}$, it is natural to define cooling schedules in terms of the dimensionless parameter $\gamma(t)$. The algebraic cooling protocol~(\ref{eq:alg-cooling}) corresponds to a physical temperature which vanishes in such a way that its inverse diverges logarithmically $T(t) \approx 4J/\alpha |\log(\tau_Q - t)|$ near~$t = \tau_Q$. To investigate the annealing dynamics with a linearly decreasing temperature $T(t) = T_0 (1-t/\tau_Q)$, we consider cooling procedures where the $\gamma$ parameter is described by an exponential function~\cite{Krapivsky2010}
\begin{eqnarray}
	1 - \gamma(t) = B \exp\left\{ - \frac{b}{(1-t/\tau_Q)^\alpha}\right\},
\end{eqnarray}
where $b, \alpha > 0$ are positive parameters and $B = \exp(b)$ is a normalization factor ensuring $\gamma(0) = 0$ and $\gamma(\tau_Q) = 1$. The case of a linearly decreasing temperature corresponds to $\alpha = 1$.

The monopole density as a function of time is shown in FIG.~\ref{fig:nm-vs-t-exp} for exponential cooling schedule with $\alpha = 1$ and~3. The dynamical evolution is again well captured by the rate equations. Similar to the case of algebraic cooling, the relaxation of magnetic monopoles is characterized by a slow decay for most of the cooling schedule, followed by an abrupt drop at the late stage. Yet, the relaxation shows a slight deceleration roughly after the freeze-out time scale, to be discussed below. This late-stage slowdown is particularly prominent in the case of~$\alpha = 3$.

The residual monopole density at $t = \tau_Q$ again exhibits a power-law dependence on the cooling rate. Here we apply the KZM to understand this scaling relation. First, we substitute $\gamma(t)$ of the exponential cooling procedure into the KZ condition~(\ref{eq:freeze-t}), the resultant transcendental equation $\hat{t} = \exp\{\tanh^{-1}[1 -B \exp(-b/(\hat{t}/\tau_Q)^\alpha) ] \}$ in the slow cooling limit can be simplified to give a freeze-out time $\hat{t} \sim \tau_Q (\ln\tau_Q)^{-1/\alpha}$. Using the scaling relation $n_m(\tau_Q) \sim \hat{\tau}^{-1} \sim \hat{t}^{-1}$ discussed previously, we obtain a universal $1/\tau_Q$ power-law relation for the residual monopole density with a logarithmic correction that depends on the parameter $\alpha$:
\begin{eqnarray}
	\label{eq:nm-left-exp}
	n_m(\tau_Q) \sim \tau_Q^{-1} \ln(\tau_Q)^{1/\alpha}.
\end{eqnarray}
As shown in FIG.~\ref{fig:nm-vs-t-exp}, the numerical results agree reasonably well with this KZM prediction.

\subsection{Effect of long-range dipolar interaction}

\noindent In pyrochlore spin-ice compounds, such as Dy$_2$Ti$_2$O$_7$ and Ho$_2$Ti$_2$O$_7$, the rare-earth ions carry a moment of 10 Bohr magnetons, $\mu \approx 10 \mu_B$. Long-range dipolar interaction plays a role of equal significance to the nearest-neighbor exchange. As discussed above, the dipolar interaction contributes to the effective nearest-neighbor coupling $J$ between the Ising spins. The dipolar term is expected to slightly modify the activation energy of monopoles $\Delta E_m$. Yet, the long-range Coulomb interaction between magnetic monopoles enhances the critical slowing down~\cite{Snyder04,Jaubert_2009}. This enhancement can be attributed to the formation of locally bound pairs of monopoles, which hinders their diffusive motion~\cite{Jaubert_2009}. As a result, the Arrhenius law cannot account for the entire low temperature relaxation time, including the intermediate quasi-plateau region (below 12~K) and the sharp upturn below 2~K. Nonetheless, the rapid increase of the relaxation time at very low temperatures, which is most relevant for the freezing in KZ scenario, can be approximated by a single exponential $\tau(T) = \tau_0 \exp(\Delta \varepsilon / T)$ with an effective barrier energy $\Delta \varepsilon$. 

For convenience, we introduce a dimensionless parameter $\lambda = \Delta E_m / \Delta \varepsilon$. The enhanced critical slowdown indicates $\lambda < 1$. The equilibrium monopole density is then $n_m \sim \xi^{-D} \sim \tau^{- \lambda}$, which implies an effective dynamical exponent is $z = D/ \lambda$. Here we consider the effects of dipolar interaction in the case of algebraic cooling schedule Eq.~(\ref{eq:alg-cooling}). Using the KZ condition~(\ref{eq:freeze-t}) to determine the freeze-out time $\hat{t}$ and the corresponding relaxation time $\hat{\tau}$, the residual monopole density is found to follow a modified scaling relation
\begin{eqnarray}
	n_m(\tau_Q) \sim \tau_Q^{-\alpha \lambda/(\alpha + 2 \lambda)}.
\end{eqnarray}
Although the correction caused by the dipolar interaction can be verified using the Glauber dynamics of Ising spins, large-scale simulations would be rather difficult due to the long-range dipolar term. A more feasible approach is to perform quench dynamics of a Coulomb gas of magnetic monopoles moving in a network of Dirac strings on the diamond lattice~\cite{Jaubert_2009} and will be left for further study.

\subsection{DISCUSSION AND OUTLOOK}

\noindent A closely related system is the 2D kagome spin ice where the Ising spins reside on a network of corner-sharing triangles~\cite{Wills02,Moller09,Chern11}. Since there are three spins in a basic triangle simplex, the ground-state manifold is governed by the 3-in-1-out or 1-in-3-out pseudo-ice rules, giving rise to a non-zero magnetic charge at every triangle. Elementary excitations, corresponding to 3-in or 3-out triangles, are not topological since they can decay into the minimum charge state by shedding the extra charge to its neighbor. The charge defects in kagome are similar to the double monopoles in pyrochlore spin ice. Moreover, while spins in the low-temperature ice phase are characterized by strong correlation, there is no emergent critical point at $T = 0$. As a result, for general cooling schedules, the residual charge defects exhibit a non-power-law dependence on the cooling rate~\cite{Libal20,Fan23}. 

The KZ mechanism has previously been investigated in artificial colloidal version of the 2D spin ice with optical traps arranged in a square lattice~\cite{Libal20}. Contrary to the ideal 2D checkerboard spin ice, the planar geometry breaks the degeneracy of the six ice-rule-obeying vertices, leading to a long-range order with staggered arrangement of the two lower-energy symmetric 2-in-2-out vertices. Although a power-law behavior of defect vertices was observed in the Langevin dynamics simulation, the obtained exponent is inconsistent with the prediction of KZM for the expected 2D Ising universality class. We believe the discrepancy could be attributed to the fact that charge defects, such as magnetic monopoles, are not necessarily associated with the Ising ordering as demonstrated in our work. On the other hand, It has been shown that the field-induced liquid-gas transition of magnetic monopoles in pyrochlore spin ice exhibits a dynamical KZ scaling of the 3D Ising universality class~\cite{Hamp15}.

Our results have firmly established the universal nonequilibrium generation of magnetic monopoles spin ices under slow cooling. Despite the absence of broken symmetries at low temperatures, pyrochlore spin ice and its 2D counterpart exhibit an unconventional critical point at $T_c = 0$. The correlation length of the highly correlated ice phase at low temperatures is controlled by emergent magnetic monopoles, which are topological defects that violate the two-in-two-out ice rules. Universal scaling relations of residual monopoles predicted by the Kibble-Zurek mechanism are confirmed by Glauber dynamics simulations as well as reaction kinetic theory. Our work opens a new avenue to the study of universal annealing dynamics of topological defects in other highly constrained systems.

\smallskip

\section{Appendix A: Relaxation time of spin ice}
\label{app:relaxation}

\begin{figure}
    \centering
    \includegraphics[width=0.95\linewidth]  {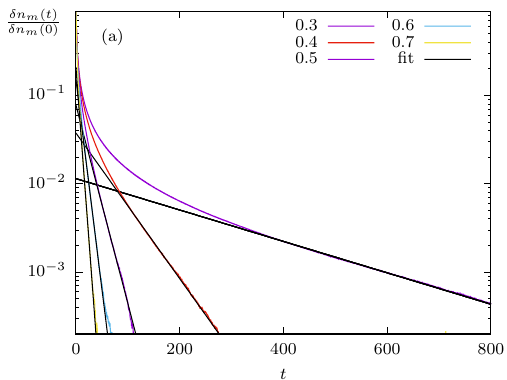}
    \includegraphics[width=0.95\linewidth]  {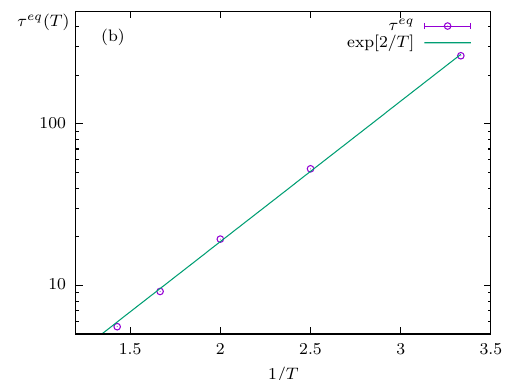}
    \caption{(a) time evolution of monopole density $n_m$ when quenched from infinite temperature to finite $T$. The results are obtained from the Glauber dynamics simulations. The density is shifted by the equilibrium density $n_{m}^{\rm eq}\left(T\right)$ and normalized for clarity. The monopole density exhibits an exponential decay that can be fitted by $a \exp\left(-t / \tau \right)$. (b) extracted relaxation time $\tau$ as a function of $1/T$. The fitting indicates that $\tau\left(T\right)\approx 0.338 J \exp{\left(2.00(3) /T\right)}$. }
    \label{fig:pyro_Q2_relax}
\end{figure}

\noindent We employ the Glauber dynamics method to simulate instant thermal quench of nearest-neighbor pyrochlore spin ice. The relaxation time of the system can be obtained from the decay of magnetic monopoles after the quench. The simulated system consists of $10^3$ cubic unit cells with $N = 16 \times 10^3$ Ising spins. All data points are averaged over 8000 randomly generated initial configurations. We quench the system from infinite temperature to a low temperature $T < J$ at time $t = 0$. 
The averaged monopole density $n_m(t)$ as a function of time is shown in FIG.~\ref{fig:pyro_Q2_relax}(a) for various final temperatures. At large $t$, the time evolution can be well approximated by an exponential decay
\begin{eqnarray}
	\delta n_m(t) = \delta n_m(0) \exp[-t/\tau(T)],
\end{eqnarray}
where $\delta n_m = n_m - n^{\rm eq}_m(T)$ is the density of excess monopoles, and $\tau(T)$ is a temperature-dependent relaxation time. The extracted relaxation time is shown in FIG.~\ref{fig:pyro_Q2_relax}(b) as a function of the inverse temperature. The agreement with the straight line, corresponding to $0.338\exp{\left(2.00(3) J /T\right)}$, in the semi-log plot shows that the relaxation time can be well approximated by an Arrhenius law with a barrier energy of $2 J$. As discussed in the main text, this result implies that the dynamical exponent $z = D$ for nearest-neighbor spin ices.

\section{Appendix B:~Reaction kinetics \& rate equations}
\label{app:reaction-kin}

\noindent 
The reaction kinetics theory in chemistry is adopted to describe the dynamical evolution of the monopoles and double monopoles in spin ices. The basic idea is to describe the time evolution in terms of the number densities of different tetrahedra in a mean-field sense. For convenience, we also borrow terms from chemical reaction theory and use species to refer to tetrahedra of different charges. In pyrochlore spin ice, there are six different species which can be classified into three types. (i) ice-rule obeying tetrahedra with zero net charge; their density is denoted as $n_0$. (ii) 3-in-1-out and 1-in-3-out tetrahedra corresponding to magnetic monopoles with charge $\pm q_m$. Their density is  denoted as $n_{\pm 1}$, respectively. (iii) all-in and all-out tetrahedra with magnetic charges $Q = \pm 2 q_m$; these are double monopoles with density $n_{\pm2}$.  

\begin{figure*}
    \includegraphics[width=0.95\linewidth]{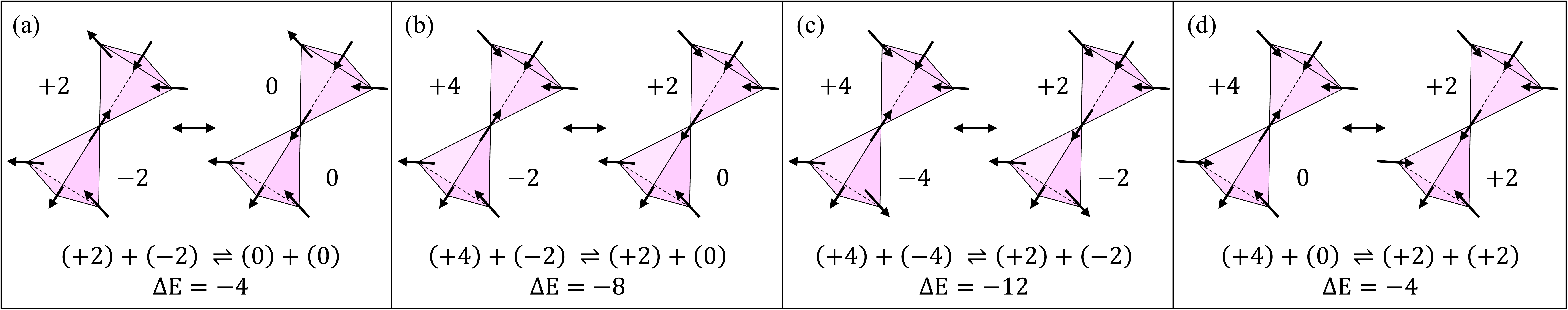}
    \caption{Different charges reactions in pyrochlore spin ice caused by a single spin-flip.  Time-reversal symmetry indicates that the reactions with all charges replaced by their opposite $Q \to -Q$ are described by exactly the same kinetics equation.}
    \label{fig:Q-reaction}
\end{figure*}

Assuming that the magnet remains spatially homogeneous during relaxation, rate equations are employed to describe the ``chemical reactions" of different tetrahedron species. The four different reactions caused by a single spin-flip are summarized in FIG.~\ref{fig:Q-reaction}. This first type, shown in FIG.~\ref{fig:Q-reaction}(a), describes the pair-annihilation and creation of magnetic monopoles $Q=\pm q_m$. The second reaction shows the annihilation of a double monopole with a single monopole of opposite charge. The third one corresponds to the conversion between a pair of monopoles and a pair of double monopoles. Finally, FIG.~\ref{fig:Q-reaction}(d) depicts the decay of a double monopole into a pair of opposite-charge fundamental monopoles. It is worth noting that single spin-flip with $\Delta E = 0$ is not listed here, as such update corresponds to the diffusive motion of fundamental monopoles.

Next we consider the transition kinetics of a general reaction
\begin{align}
	\label{eq:Q-reaction}
    Q_A + Q_B \rightleftharpoons Q_C + Q_D,
\end{align}
where $Q_A, ~Q_B$ are the initial reactants, and $Q_C, ~Q_D$ are the final products. The two-way harpoon indicates that the reaction can occur in both forward and reversed directions. We note that, as these reactions are due to a flipping of magnetic dipole, the total charge is conserved. 
It is convenient to choose the forward direction as the one that lowers the total energy, i.e., $\Delta E < 0$.
In other words, the forward reaction is the decay or the annihilation of magnetic charges, while the reversed reaction is the excitation of magnetic charges.
The rate of a reaction is proportional to the densities of the reactants. For example, the transition rate of forward reaction for Eq.~(\ref{eq:Q-reaction}) is $v_+ \propto n_{Q_A} n_{Q_B}$. The net rate of reaction in the forward direction is then
\begin{align}
    v =  k_+ n_{Q_A} n_{Q_B} - k_{-1}n_{Q_C} n_{Q_D},
\end{align}
where $n_Q$ is the density of tetrahedron with charge $Q$, and $k_{\pm}$ denote the reaction coefficients of forward/reversed reactions, respectively. 
These reaction coefficients, however, are not independent. When the system reaches equilibrium, the net change is zero $v = 0$, which in turn means $k_+/k_- = n^{\rm eq}_{Q_C} n^{\rm eq}_{Q_D} / n^{\rm eq}_{Q_A} n^{\rm eq}_{Q_B}$.
The equilibrium densities of the various species are given by the Boltzmann distribution, $n^{\rm eq}_Q = g_Q \,e^{-\beta E_Q} / Z$, where $Z$ is the partition function, $E_Q$ is the energy of charge species $Q$, and $g_Q$ is its degeneracy. We thus have
\begin{eqnarray}
	\label{eq:ratio_k}
	\frac{k_+}{k_-} =  \frac{g_{Q_C} g_{Q_D} }{g_{Q_A} g_{Q_B}}e^{-\beta \Delta E},
\end{eqnarray}
where $\Delta E$ is the energy difference between products and reactants. In general, the reaction coefficients $k_\pm$ can be expressed as
\begin{eqnarray}
	\label{eq:e_av}
	k_{\pm} = A_{\pm} e^{-\beta \varepsilon_{\pm}},
\end{eqnarray}
where $\varepsilon_{\pm}$ are the activation energies for the forward/backward reactions, respectively. In chemical reactions that often involve an intermediate state, these energy barriers are the energy differences between the intermediate state and the initial/final state, respectively. The coefficients $A_{\pm}$ are now nearly temperature independent. Let $E^*$ be the energy of the intermediate state, we have $\varepsilon_+ = E^* - (E_{Q_A} + E_{Q_B})$ and $\varepsilon_- = E^* - (E_{Q_C} + E_{Q_D})$. Substitute Eq.~(\ref{eq:e_av}) into the ration in Eq.~(\ref{eq:ratio_k}), and using the fact that $\varepsilon_+ - \varepsilon_- = \Delta E$, we obtain the ratio between the two pre-factors
\begin{eqnarray}
	\frac{A_+}{A_-} =  \frac{g_{Q_C} g_{Q_D} }{g_{Q_A} g_{Q_B}}.
\end{eqnarray} 
The overall reaction rate, and in particular its temperature dependence, naturally also depends on the energy level $E^*$ of the intermediate state. However, for Ising spins with Glauber dynamics, the transition rate only depends on the energy difference $\Delta E$, which does not involve any intermediate state. Or equivalently, the initial state with higher energy serves as such intermediate, hence $\varepsilon_+=0$ and $\varepsilon_- = |\Delta E|$.

With these simplifications, there is only one independent parameter, e.g., $A_-$, for the determination of the net reaction rate
\begin{align}
    v = A_{-} \left( \frac{g_{Q_C} g_{Q_D}}{g_{Q_A} g_{Q_B}} n_{Q_A} n_{Q_B} - e^{-\beta \left|\Delta E\right|} n_{Q_C} n_{Q_D} \right).
    \label{eq::spin_ice_reaction_rate}
\end{align}
When a charged species is involved in multiple reactions, the rate equation of its density should include contributions from all possible reactions. 

For charge species involved in multiple reactions simultaneously, their rate equation should include the contributions of every reaction
\begin{align}
    \frac{dn_Q}{dt} = \sum_m \left(r_{m,Q} - s_{m,Q} \right) v_m,
    \label{eq::rate_equation}
\end{align}
where $v_m$ is the rate of the $m$-th reaction, $r_{m,Q}$ and $s_{m,Q}$ are the stoichiometric coefficients species-$Q$ in the reactants and products, respectively, of the $m$-th reaction. 

To further simplify the rate equation, we utilize the charge symmetry of the spin ice system, assume that the charge densities of species with opposite signs are the same, and define defect densities of the system, $n_m = (n_{+1} + n_{-1})$ and $n_{2m} = (n_{+2} + n_{-2})$. And the density of background tetrahedra satisfying the ice rules is $n_0 = 1 - n_m - n_{2m}$.
Based on the possible reactions and the properties of charges in pyrochlore spin ice, we can see that the densities of charge defects satisfy the following ordinary differential equations,
\begin{align}
    \label{eq:rate_eq1}
    &  \frac{dn_{m}}{dt} = A_1\! \left(2 e^{-4\beta J} n_0^2 - \frac{9}{8} n_{m}^2 \right)     \\
                   & \,\,     - A_3\! \left(\frac{1}{2} e^{-12\beta J} n_{m}^2 - 8 n_{2m}^2 \right)   
                     - A_4 \!\left(e^{-4\beta J} n_{m}^2 - \frac{16}{3} n_0 n_{2m} \right) \nonumber
\end{align}
\begin{align} 
	\label{eq:rate_eq2}  
    &   \frac{dn_{2m}}{dt} = A_2 \!\left(e^{-8\beta J} n_0 n_{m} - 3 n_{m} n_{2m} \right)  \\
    & \,\, + A_3\! \left(\frac{1}{2} e^{-12\beta J} n_{m}^2 - 8 n_{2m}^2 \right)  
       + A_4\! \left(\frac{1}{2}e^{-4\beta J} n_{m}^2 - \frac{8}{3} n_0 n_{2m} \right) \nonumber
\end{align}
where  $\beta(T) = 1/T(t)$ is the time-dependent inverse temperature. The four coefficients, $A_1, \cdots, A_4$ describe the overall reaction rates of the four reaction processes in FIG.~\ref{fig:Q-reaction}(a)--(d), and are obtained by fitting with the Glauber dynamics simulations. For the quench simulations, the initially random spins at $T = \infty$ corresponds to initial conditions $n_m(0)=\frac{1}{2}$ and $n_{2m}(0)=\frac{1}{8}$. 

The rate equation~(\ref{eq:rate-nm}) for magnetic monopoles at low temperatures is obtained from Eq.~(\ref{eq:rate_eq1}) using the approximations $n_0 = 1-n_m - n_{2m} \approx 1$. The various coefficients there are $\mathcal{A}_0 = 2 A_1 e^{-4\beta}$, $\mathcal{A}_1 = 16 A_4/3$, $\mathcal{A}_2 = 8 A_3$, and $\mathcal{B} = 9A_1/8 + A_3 e^{-12\beta J}/2 + A_4 e^{-4\beta J}$.

\subsection{Appendix C: Asymptotic solution for residual double monopoles}
\label{app:asymptotic}

\noindent At low temperatures after the freeze-out time, the rate equation for double monopole is dominated by the $A_4$ term, corresponding to the reaction shown in FIG.~\ref{fig:Q-reaction}(d). This is justified mathematically by the fact that $e^{-4\beta J} \gg e^{-8 \beta J} \gg e^{-12\beta J}$ for the source term, and $n_0 \gg n_m \gg n_{2m}$. The resultant rate equation is shown in Eq.~(\ref{eq:rate-n2m}) in the main text with the intrinsic lifetime of double monopole given by $\tau_{2m} = 3/(8A_4)$. For convenience, let $t_* = \tau_Q - \hat{t}$ be the freeze-out moment during the cooling. The monopole density remains roughly a constant $\hat{n}_m = n_m(t_*) \approx  n_m(\tau_Q)$ for time interval $t_* < t < \tau_Q$. The rate equation then becomes
\begin{eqnarray}
	\frac{dn_{2m}}{dt} = \frac{3 \hat{n}_m^2}{16 \tau_{2m}} e^{-4\beta(t) J}  - \frac{n_{2m} }{ \tau_{2m}},
\end{eqnarray}
Integrating this equation from $t_*$ to $t$ gives 
\begin{align}
	& n_{2m}(t) = n_{2m}(t_*) e^{-(t-t_*)/\tau_{2m}} \\
	& \qquad  + \frac{3 \hat{n}_m^2}{16 \tau_{2m}} \int_{t_*}^t e^{-4\beta(s)} e^{-(t-s)/\tau_{2m}} ds. \nonumber
\end{align}
Here the density of double monopole $n_{2m}(t_*)$ at the freeze-out instant is given in Eq.~(\ref{eq:n2m-freeze}). Assuming a fast decay of double monopoles $\tau_{2m} \ll \hat{t}$, the exponential factor of the first term at $t = \tau_Q$ is then negligible $e^{-\hat{t}/\tau_{2m}} \ll 1$. To evaluate the remaining integral, we introduce a change of variable $\eta = (\tau_Q - s)/\tau_{2m}$ and express the Arrhenius factor $e^{-4\beta J}$ in terms of the dimensionless $\gamma$. The residual density at the end of cooling becomes
\begin{eqnarray}
	n_{2m}(\tau_Q) = \frac{3 \hat{n}_m^2}{16} \int_{0}^{\hat{t} / \tau_{2m}} \frac{1-\gamma(\eta)}{1+\gamma(\eta)} e^{-\eta} d\eta.
\end{eqnarray}
In the low temperature regime at the end of cooling, $\gamma \approx 1$, we can approximate the denominator by $1+\gamma \approx 2$. Substituting the algebraic cooling protocol~(\ref{eq:alg-cooling}) for $1-\gamma$, we obtain
\begin{eqnarray}
	n_{2m}(\tau_Q) = \frac{3 A \hat{n}_m^2 }{16} \left(\frac{\tau_{2m}}{\tau_Q} \right)^\alpha \int_0^{\hat{t}/\tau_{2m}} \eta^\alpha e^{-\eta} d\eta.
\end{eqnarray}
The remaining integral can be readily evaluated in the $\tau_{2m} \gg \hat{t}$ limit:
\begin{eqnarray}
	n_{2m}(\tau_Q) = \frac{3 A \Gamma(1+\alpha) \hat{n}_m^2 }{16} \left(\frac{\tau_{2m}}{\tau_Q} \right)^\alpha.
\end{eqnarray}
where $\Gamma(x)$ is the Gamma function. Using scaling relation~(\ref{eq:KZM-nm}) for the residual monopole density $\hat{n}_m$, the above equation leads to the power-law behavior in Eq.~(\ref{eq:n2m-left}).

\bigskip
\bigskip

\begin{acknowledgments}
{\bf Acknowledgments.} GWC is partially supported by the US Department of Energy Basic Energy Sciences under Contract No. DE-SC0020330. The authors also acknowledge the support of Research Computing at the University of Virginia. 
\end{acknowledgments}

\bibliography{SpinIceRef}

\begin{thebibliography}{78}%
\makeatletter
\providecommand \@ifxundefined [1]{%
 \@ifx{#1\undefined}
}%
\providecommand \@ifnum [1]{%
 \ifnum #1\expandafter \@firstoftwo
 \else \expandafter \@secondoftwo
 \fi
}%
\providecommand \@ifx [1]{%
 \ifx #1\expandafter \@firstoftwo
 \else \expandafter \@secondoftwo
 \fi
}%
\providecommand \natexlab [1]{#1}%
\providecommand \enquote  [1]{``#1''}%
\providecommand \bibnamefont  [1]{#1}%
\providecommand \bibfnamefont [1]{#1}%
\providecommand \citenamefont [1]{#1}%
\providecommand \href@noop [0]{\@secondoftwo}%
\providecommand \href [0]{\begingroup \@sanitize@url \@href}%
\providecommand \@href[1]{\@@startlink{#1}\@@href}%
\providecommand \@@href[1]{\endgroup#1\@@endlink}%
\providecommand \@sanitize@url [0]{\catcode `\\12\catcode `\$12\catcode
  `\&12\catcode `\#12\catcode `\^12\catcode `\_12\catcode `\%12\relax}%
\providecommand \@@startlink[1]{}%
\providecommand \@@endlink[0]{}%
\providecommand \url  [0]{\begingroup\@sanitize@url \@url }%
\providecommand \@url [1]{\endgroup\@href {#1}{\urlprefix }}%
\providecommand \urlprefix  [0]{URL }%
\providecommand \Eprint [0]{\href }%
\providecommand \doibase [0]{https://doi.org/}%
\providecommand \selectlanguage [0]{\@gobble}%
\providecommand \bibinfo  [0]{\@secondoftwo}%
\providecommand \bibfield  [0]{\@secondoftwo}%
\providecommand \translation [1]{[#1]}%
\providecommand \BibitemOpen [0]{}%
\providecommand \bibitemStop [0]{}%
\providecommand \bibitemNoStop [0]{.\EOS\space}%
\providecommand \EOS [0]{\spacefactor3000\relax}%
\providecommand \BibitemShut  [1]{\csname bibitem#1\endcsname}%
\let\auto@bib@innerbib\@empty
%</preamble>
\bibitem [{\citenamefont {Kibble}(1976)}]{Kibble76a}%
  \BibitemOpen
  \bibfield  {author} {\bibinfo {author} {\bibfnamefont {T.~W.~B.}\
  \bibnamefont {Kibble}},\ }\bibfield  {title} {\bibinfo {title} {Topology of
  cosmic domains and strings},\ }\href
  {http://stacks.iop.org/0305-4470/9/i=8/a=029} {\bibfield  {journal} {\bibinfo
   {journal} {\href{http://stacks.iop.org/0305-4470/9/i=8/a=029}{J. of Phys. A:
  Math. Gen.}}\ }\textbf {\bibinfo {volume} {9}},\ \bibinfo {pages} {1387}
  (\bibinfo {year} {1976})}\BibitemShut {NoStop}%
\bibitem [{\citenamefont {Kibble}(1980)}]{Kibble76b}%
  \BibitemOpen
  \bibfield  {author} {\bibinfo {author} {\bibfnamefont {T.~W.~B.}\
  \bibnamefont {Kibble}},\ }\bibfield  {title} {\bibinfo {title} {Some
  implications of a cosmological phase transition},\ }\href
  {http://www.sciencedirect.com/science/article/pii/0370157380900915}
  {\bibfield  {journal} {\bibinfo  {journal}
  {\href{http://www.sciencedirect.com/science/article/pii/0370157380900915}{Phys.
  Reports}}\ }\textbf {\bibinfo {volume} {67}},\ \bibinfo {pages} {183}
  (\bibinfo {year} {1980})}\BibitemShut {NoStop}%
\bibitem [{\citenamefont {Zurek}(1985)}]{Zurek96a}%
  \BibitemOpen
  \bibfield  {author} {\bibinfo {author} {\bibfnamefont {W.~H.}\ \bibnamefont
  {Zurek}},\ }\bibfield  {title} {\bibinfo {title} {Cosmological experiments in
  superfluid helium?},\ }\href {http://dx.doi.org/10.1038/317505a0} {\bibfield
  {journal} {\bibinfo  {journal}
  {\href{http://dx.doi.org/10.1038/317505a0}{Nature}}\ }\textbf {\bibinfo
  {volume} {317}},\ \bibinfo {pages} {505} (\bibinfo {year}
  {1985})}\BibitemShut {NoStop}%
\bibitem [{\citenamefont {Zurek}(1993)}]{Zurek96c}%
  \BibitemOpen
  \bibfield  {author} {\bibinfo {author} {\bibfnamefont {W.~H.}\ \bibnamefont
  {Zurek}},\ }\bibfield  {title} {\bibinfo {title} {Cosmological experiments in
  condensed matter systems},\ }\href
  {http://www.sciencedirect.com/science/article/pii/S0370157396000099}
  {\bibfield  {journal} {\bibinfo  {journal}
  {\href{http://www.sciencedirect.com/science/article/pii/S0370157396000099}{Phys.
  Reports}}\ }\textbf {\bibinfo {volume} {276}},\ \bibinfo {pages} {177}
  (\bibinfo {year} {1993})}\BibitemShut {NoStop}%
\bibitem [{\citenamefont {del Campo}\ and\ \citenamefont {Zurek}(2014)}]{DZ14}%
  \BibitemOpen
  \bibfield  {author} {\bibinfo {author} {\bibfnamefont {A.}~\bibnamefont {del
  Campo}}\ and\ \bibinfo {author} {\bibfnamefont {W.~H.}\ \bibnamefont
  {Zurek}},\ }\bibfield  {title} {\bibinfo {title} {Universality of phase
  transition dynamics: Topological defects from symmetry breaking},\ }\href
  {https://doi.org/10.1142/S0217751X1430018X} {\bibfield  {journal} {\bibinfo
  {journal} {Int. J. Mod. Phys. A}\ }\textbf {\bibinfo {volume} {29}},\
  \bibinfo {pages} {1430018} (\bibinfo {year} {2014})}\BibitemShut {NoStop}%
\bibitem [{\citenamefont {Zeldovich}\ and\ \citenamefont
  {Khlopov}(1978)}]{Zeldovich78}%
  \BibitemOpen
  \bibfield  {author} {\bibinfo {author} {\bibfnamefont {Y.}~\bibnamefont
  {Zeldovich}}\ and\ \bibinfo {author} {\bibfnamefont {M.}~\bibnamefont
  {Khlopov}},\ }\bibfield  {title} {\bibinfo {title} {On the concentration of
  relic magnetic monopoles in the universe},\ }\href
  {https://doi.org/https://doi.org/10.1016/0370-2693(78)90232-0} {\bibfield
  {journal} {\bibinfo  {journal} {Physics Letters B}\ }\textbf {\bibinfo
  {volume} {79}},\ \bibinfo {pages} {239} (\bibinfo {year} {1978})}\BibitemShut
  {NoStop}%
\bibitem [{\citenamefont {Preskill}(1979)}]{Preskill79}%
  \BibitemOpen
  \bibfield  {author} {\bibinfo {author} {\bibfnamefont {J.~P.}\ \bibnamefont
  {Preskill}},\ }\bibfield  {title} {\bibinfo {title} {Cosmological production
  of superheavy magnetic monopoles},\ }\href
  {https://doi.org/10.1103/PhysRevLett.43.1365} {\bibfield  {journal} {\bibinfo
   {journal} {Phys. Rev. Lett.}\ }\textbf {\bibinfo {volume} {43}},\ \bibinfo
  {pages} {1365} (\bibinfo {year} {1979})}\BibitemShut {NoStop}%
\bibitem [{\citenamefont {Einhorn}\ \emph {et~al.}(1980)\citenamefont
  {Einhorn}, \citenamefont {Stein},\ and\ \citenamefont
  {Toussaint}}]{Einhorn80}%
  \BibitemOpen
  \bibfield  {author} {\bibinfo {author} {\bibfnamefont {M.~B.}\ \bibnamefont
  {Einhorn}}, \bibinfo {author} {\bibfnamefont {D.~L.}\ \bibnamefont {Stein}},\
  and\ \bibinfo {author} {\bibfnamefont {D.}~\bibnamefont {Toussaint}},\
  }\bibfield  {title} {\bibinfo {title} {Are grand unified theories compatible
  with standard cosmology?},\ }\href {https://doi.org/10.1103/PhysRevD.21.3295}
  {\bibfield  {journal} {\bibinfo  {journal} {Phys. Rev. D}\ }\textbf {\bibinfo
  {volume} {21}},\ \bibinfo {pages} {3295} (\bibinfo {year}
  {1980})}\BibitemShut {NoStop}%
\bibitem [{\citenamefont {Guth}(1981)}]{Guth81}%
  \BibitemOpen
  \bibfield  {author} {\bibinfo {author} {\bibfnamefont {A.~H.}\ \bibnamefont
  {Guth}},\ }\bibfield  {title} {\bibinfo {title} {Inflationary universe: A
  possible solution to the horizon and flatness problems},\ }\href
  {https://doi.org/10.1103/PhysRevD.23.347} {\bibfield  {journal} {\bibinfo
  {journal} {Phys. Rev. D}\ }\textbf {\bibinfo {volume} {23}},\ \bibinfo
  {pages} {347} (\bibinfo {year} {1981})}\BibitemShut {NoStop}%
\bibitem [{\citenamefont {Chandran}\ \emph {et~al.}(2012)\citenamefont
  {Chandran}, \citenamefont {Erez}, \citenamefont {Gubser},\ and\ \citenamefont
  {Sondhi}}]{Chandran12}%
  \BibitemOpen
  \bibfield  {author} {\bibinfo {author} {\bibfnamefont {A.}~\bibnamefont
  {Chandran}}, \bibinfo {author} {\bibfnamefont {A.}~\bibnamefont {Erez}},
  \bibinfo {author} {\bibfnamefont {S.~S.}\ \bibnamefont {Gubser}},\ and\
  \bibinfo {author} {\bibfnamefont {S.~L.}\ \bibnamefont {Sondhi}},\ }\bibfield
   {title} {\bibinfo {title} {Kibble-{Z}urek problem: Universality and the
  scaling limit},\ }\href {https://doi.org/10.1103/PhysRevB.86.064304}
  {\bibfield  {journal} {\bibinfo  {journal} {Phys. Rev. B}\ }\textbf {\bibinfo
  {volume} {86}},\ \bibinfo {pages} {064304} (\bibinfo {year}
  {2012})}\BibitemShut {NoStop}%
\bibitem [{\citenamefont {Nikoghosyan}\ \emph {et~al.}(2016)\citenamefont
  {Nikoghosyan}, \citenamefont {Nigmatullin},\ and\ \citenamefont
  {Plenio}}]{Nikoghosyan16}%
  \BibitemOpen
  \bibfield  {author} {\bibinfo {author} {\bibfnamefont {G.}~\bibnamefont
  {Nikoghosyan}}, \bibinfo {author} {\bibfnamefont {R.}~\bibnamefont
  {Nigmatullin}},\ and\ \bibinfo {author} {\bibfnamefont {M.~B.}\ \bibnamefont
  {Plenio}},\ }\bibfield  {title} {\bibinfo {title} {Universality in the
  dynamics of second-order phase transitions},\ }\href
  {https://doi.org/10.1103/PhysRevLett.116.080601} {\bibfield  {journal}
  {\bibinfo  {journal} {Phys. Rev. Lett.}\ }\textbf {\bibinfo {volume} {116}},\
  \bibinfo {pages} {080601} (\bibinfo {year} {2016})}\BibitemShut {NoStop}%
\bibitem [{\citenamefont {Laguna}\ and\ \citenamefont
  {Zurek}(1998)}]{Laguna98}%
  \BibitemOpen
  \bibfield  {author} {\bibinfo {author} {\bibfnamefont {P.}~\bibnamefont
  {Laguna}}\ and\ \bibinfo {author} {\bibfnamefont {W.~H.}\ \bibnamefont
  {Zurek}},\ }\bibfield  {title} {\bibinfo {title} {Critical dynamics of
  symmetry breaking: Quenches, dissipation, and cosmology},\ }\href
  {https://doi.org/10.1103/PhysRevD.58.085021} {\bibfield  {journal} {\bibinfo
  {journal} {Phys. Rev. D}\ }\textbf {\bibinfo {volume} {58}},\ \bibinfo
  {pages} {085021} (\bibinfo {year} {1998})}\BibitemShut {NoStop}%
\bibitem [{\citenamefont {Antunes}\ \emph {et~al.}(2006)\citenamefont
  {Antunes}, \citenamefont {Gandra},\ and\ \citenamefont {Rivers}}]{Antunes06}%
  \BibitemOpen
  \bibfield  {author} {\bibinfo {author} {\bibfnamefont {N.~D.}\ \bibnamefont
  {Antunes}}, \bibinfo {author} {\bibfnamefont {P.}~\bibnamefont {Gandra}},\
  and\ \bibinfo {author} {\bibfnamefont {R.~J.}\ \bibnamefont {Rivers}},\
  }\bibfield  {title} {\bibinfo {title} {Is domain formation decided before or
  after the transition?},\ }\href {https://doi.org/10.1103/PhysRevD.73.125003}
  {\bibfield  {journal} {\bibinfo  {journal} {Phys. Rev. D}\ }\textbf {\bibinfo
  {volume} {73}},\ \bibinfo {pages} {125003} (\bibinfo {year}
  {2006})}\BibitemShut {NoStop}%
\bibitem [{\citenamefont {Zurek}(2009)}]{Zurek09}%
  \BibitemOpen
  \bibfield  {author} {\bibinfo {author} {\bibfnamefont {W.~H.}\ \bibnamefont
  {Zurek}},\ }\bibfield  {title} {\bibinfo {title} {Causality in condensates:
  Gray solitons as relics of bec formation},\ }\href
  {https://doi.org/10.1103/PhysRevLett.102.105702} {\bibfield  {journal}
  {\bibinfo  {journal}
  {\href{https://link.aps.org/doi/10.1103/PhysRevLett.102.105702}{Phys. Rev.
  Lett.}}\ }\textbf {\bibinfo {volume} {102}},\ \bibinfo {pages} {105702}
  (\bibinfo {year} {2009})}\BibitemShut {NoStop}%
\bibitem [{\citenamefont {Suzuki}(2009)}]{Suzuki09a}%
  \BibitemOpen
  \bibfield  {author} {\bibinfo {author} {\bibfnamefont {S.}~\bibnamefont
  {Suzuki}},\ }\bibfield  {title} {\bibinfo {title} {Cooling dynamics of pure
  and random ising chains},\ }\href
  {http://stacks.iop.org/1742-5468/2009/i=03/a=P03032} {\bibfield  {journal}
  {\bibinfo  {journal}
  {\href{http://stacks.iop.org/1742-5468/2009/i=03/a=P03032}{J. Stat. Mech.:
  Theo. Exp.}}\ }\textbf {\bibinfo {volume} {2009}},\ \bibinfo {pages} {P03032}
  (\bibinfo {year} {2009})}\BibitemShut {NoStop}%
\bibitem [{\citenamefont {del Campo}\ \emph {et~al.}(2010)\citenamefont {del
  Campo}, \citenamefont {De~Chiara}, \citenamefont {Morigi}, \citenamefont
  {Plenio},\ and\ \citenamefont {Retzker}}]{delcampo10}%
  \BibitemOpen
  \bibfield  {author} {\bibinfo {author} {\bibfnamefont {A.}~\bibnamefont {del
  Campo}}, \bibinfo {author} {\bibfnamefont {G.}~\bibnamefont {De~Chiara}},
  \bibinfo {author} {\bibfnamefont {G.}~\bibnamefont {Morigi}}, \bibinfo
  {author} {\bibfnamefont {M.~B.}\ \bibnamefont {Plenio}},\ and\ \bibinfo
  {author} {\bibfnamefont {A.}~\bibnamefont {Retzker}},\ }\bibfield  {title}
  {\bibinfo {title} {Structural defects in ion chains by quenching the external
  potential: The inhomogeneous {K}ibble-{Z}urek mechanism},\ }\href
  {https://doi.org/10.1103/PhysRevLett.105.075701} {\bibfield  {journal}
  {\bibinfo  {journal} {Phys. Rev. Lett.}\ }\textbf {\bibinfo {volume} {105}},\
  \bibinfo {pages} {075701} (\bibinfo {year} {2010})}\BibitemShut {NoStop}%
\bibitem [{\citenamefont {Das}\ \emph {et~al.}(2012)\citenamefont {Das},
  \citenamefont {Sabbatini},\ and\ \citenamefont {Zurek}}]{Das12}%
  \BibitemOpen
  \bibfield  {author} {\bibinfo {author} {\bibfnamefont {A.}~\bibnamefont
  {Das}}, \bibinfo {author} {\bibfnamefont {J.}~\bibnamefont {Sabbatini}},\
  and\ \bibinfo {author} {\bibfnamefont {W.~H.}\ \bibnamefont {Zurek}},\
  }\bibfield  {title} {\bibinfo {title} {Winding up superfluid in a torus via
  bose einstein condensation},\ }\href {https://doi.org/10.1038/srep00352}
  {\bibfield  {journal} {\bibinfo  {journal} {Sci. Rep.}\ }\textbf {\bibinfo
  {volume} {2}},\ \bibinfo {pages} {352} (\bibinfo {year} {2012})}\BibitemShut
  {NoStop}%
\bibitem [{\citenamefont {Sonner}\ \emph {et~al.}(2015)\citenamefont {Sonner},
  \citenamefont {del Campo},\ and\ \citenamefont {Zurek}}]{Sonner15}%
  \BibitemOpen
  \bibfield  {author} {\bibinfo {author} {\bibfnamefont {J.}~\bibnamefont
  {Sonner}}, \bibinfo {author} {\bibfnamefont {A.}~\bibnamefont {del Campo}},\
  and\ \bibinfo {author} {\bibfnamefont {W.~H.}\ \bibnamefont {Zurek}},\
  }\bibfield  {title} {\bibinfo {title} {Universal far-from-equilibrium
  dynamics of a holographic superconductor},\ }\href
  {https://doi.org/10.1038/ncomms8406} {\bibfield  {journal} {\bibinfo
  {journal} {Nat. Comm.}\ }\textbf {\bibinfo {volume} {6}},\ \bibinfo {pages}
  {7406} (\bibinfo {year} {2015})}\BibitemShut {NoStop}%
\bibitem [{\citenamefont {G\'omez-Ruiz}\ \emph {et~al.}(2020)\citenamefont
  {G\'omez-Ruiz}, \citenamefont {Mayo},\ and\ \citenamefont {del
  Campo}}]{GRMdC19}%
  \BibitemOpen
  \bibfield  {author} {\bibinfo {author} {\bibfnamefont {F.~J.}\ \bibnamefont
  {G\'omez-Ruiz}}, \bibinfo {author} {\bibfnamefont {J.~J.}\ \bibnamefont
  {Mayo}},\ and\ \bibinfo {author} {\bibfnamefont {A.}~\bibnamefont {del
  Campo}},\ }\bibfield  {title} {\bibinfo {title} {Full counting statistics of
  topological defects after crossing a phase transition},\ }\href
  {https://doi.org/10.1103/PhysRevLett.124.240602} {\bibfield  {journal}
  {\bibinfo  {journal} {Phys. Rev. Lett.}\ }\textbf {\bibinfo {volume} {124}},\
  \bibinfo {pages} {240602} (\bibinfo {year} {2020})}\BibitemShut {NoStop}%
\bibitem [{\citenamefont {Xu}\ \emph {et~al.}(2014)\citenamefont {Xu},
  \citenamefont {Han}, \citenamefont {Sun}, \citenamefont {Xu}, \citenamefont
  {Tang}, \citenamefont {Li},\ and\ \citenamefont {Guo}}]{Guo14}%
  \BibitemOpen
  \bibfield  {author} {\bibinfo {author} {\bibfnamefont {X.-Y.}\ \bibnamefont
  {Xu}}, \bibinfo {author} {\bibfnamefont {Y.-J.}\ \bibnamefont {Han}},
  \bibinfo {author} {\bibfnamefont {K.}~\bibnamefont {Sun}}, \bibinfo {author}
  {\bibfnamefont {J.-S.}\ \bibnamefont {Xu}}, \bibinfo {author} {\bibfnamefont
  {J.-S.}\ \bibnamefont {Tang}}, \bibinfo {author} {\bibfnamefont {C.-F.}\
  \bibnamefont {Li}},\ and\ \bibinfo {author} {\bibfnamefont {G.-C.}\
  \bibnamefont {Guo}},\ }\bibfield  {title} {\bibinfo {title} {Quantum
  simulation of {L}andau-{Z}ener model dynamics supporting the {K}ibble-{Z}urek
  mechanism},\ }\href {https://doi.org/10.1103/PhysRevLett.112.035701}
  {\bibfield  {journal} {\bibinfo  {journal} {Phys. Rev. Lett.}\ }\textbf
  {\bibinfo {volume} {112}},\ \bibinfo {pages} {035701} (\bibinfo {year}
  {2014})}\BibitemShut {NoStop}%
\bibitem [{\citenamefont {Wang}\ \emph {et~al.}(2014)\citenamefont {Wang},
  \citenamefont {Zhou}, \citenamefont {Tu}, \citenamefont {Jiang},
  \citenamefont {Guo},\ and\ \citenamefont {Guo}}]{Wang14}%
  \BibitemOpen
  \bibfield  {author} {\bibinfo {author} {\bibfnamefont {L.}~\bibnamefont
  {Wang}}, \bibinfo {author} {\bibfnamefont {C.}~\bibnamefont {Zhou}}, \bibinfo
  {author} {\bibfnamefont {T.}~\bibnamefont {Tu}}, \bibinfo {author}
  {\bibfnamefont {H.-W.}\ \bibnamefont {Jiang}}, \bibinfo {author}
  {\bibfnamefont {G.-P.}\ \bibnamefont {Guo}},\ and\ \bibinfo {author}
  {\bibfnamefont {G.-C.}\ \bibnamefont {Guo}},\ }\bibfield  {title} {\bibinfo
  {title} {Quantum simulation of the {K}ibble-{Z}urek mechanism using a
  semiconductor electron charge qubit},\ }\href
  {https://doi.org/10.1103/PhysRevA.89.022337} {\bibfield  {journal} {\bibinfo
  {journal} {Phys. Rev. A}\ }\textbf {\bibinfo {volume} {89}},\ \bibinfo
  {pages} {022337} (\bibinfo {year} {2014})}\BibitemShut {NoStop}%
\bibitem [{\citenamefont {Gong}\ \emph {et~al.}(2016)\citenamefont {Gong},
  \citenamefont {Wen}, \citenamefont {Sun}, \citenamefont {Zhang},
  \citenamefont {Lan}, \citenamefont {Zhou}, \citenamefont {Fan}, \citenamefont
  {Liu}, \citenamefont {Tan}, \citenamefont {Yu}, \citenamefont {Yu},
  \citenamefont {Zhu}, \citenamefont {Han},\ and\ \citenamefont {Wu}}]{Wu16}%
  \BibitemOpen
  \bibfield  {author} {\bibinfo {author} {\bibfnamefont {M.}~\bibnamefont
  {Gong}}, \bibinfo {author} {\bibfnamefont {X.}~\bibnamefont {Wen}}, \bibinfo
  {author} {\bibfnamefont {G.}~\bibnamefont {Sun}}, \bibinfo {author}
  {\bibfnamefont {D.-W.}\ \bibnamefont {Zhang}}, \bibinfo {author}
  {\bibfnamefont {D.}~\bibnamefont {Lan}}, \bibinfo {author} {\bibfnamefont
  {Y.}~\bibnamefont {Zhou}}, \bibinfo {author} {\bibfnamefont {Y.}~\bibnamefont
  {Fan}}, \bibinfo {author} {\bibfnamefont {Y.}~\bibnamefont {Liu}}, \bibinfo
  {author} {\bibfnamefont {X.}~\bibnamefont {Tan}}, \bibinfo {author}
  {\bibfnamefont {H.}~\bibnamefont {Yu}}, \bibinfo {author} {\bibfnamefont
  {Y.}~\bibnamefont {Yu}}, \bibinfo {author} {\bibfnamefont {S.-L.}\
  \bibnamefont {Zhu}}, \bibinfo {author} {\bibfnamefont {S.}~\bibnamefont
  {Han}},\ and\ \bibinfo {author} {\bibfnamefont {P.}~\bibnamefont {Wu}},\
  }\bibfield  {title} {\bibinfo {title} {Simulating the {K}ibble-{Z}urek
  mechanism of the {I}sing model with a superconducting qubit system},\ }\href
  {https://doi.org/10.1038/srep22667} {\bibfield  {journal} {\bibinfo
  {journal} {Sci Rep.}\ }\textbf {\bibinfo {volume} {6}},\ \bibinfo {pages}
  {22667} (\bibinfo {year} {2016})}\BibitemShut {NoStop}%
\bibitem [{\citenamefont {Cui}\ \emph {et~al.}(2016)\citenamefont {Cui},
  \citenamefont {Huang}, \citenamefont {Wang}, \citenamefont {Cao},
  \citenamefont {Wang}, \citenamefont {Lv}, \citenamefont {Luo}, \citenamefont
  {del Campo}, \citenamefont {Han}, \citenamefont {Li},\ and\ \citenamefont
  {Guo}}]{Cui16}%
  \BibitemOpen
  \bibfield  {author} {\bibinfo {author} {\bibfnamefont {J.-M.}\ \bibnamefont
  {Cui}}, \bibinfo {author} {\bibfnamefont {Y.-F.}\ \bibnamefont {Huang}},
  \bibinfo {author} {\bibfnamefont {Z.}~\bibnamefont {Wang}}, \bibinfo {author}
  {\bibfnamefont {D.-Y.}\ \bibnamefont {Cao}}, \bibinfo {author} {\bibfnamefont
  {J.}~\bibnamefont {Wang}}, \bibinfo {author} {\bibfnamefont {W.-M.}\
  \bibnamefont {Lv}}, \bibinfo {author} {\bibfnamefont {L.}~\bibnamefont
  {Luo}}, \bibinfo {author} {\bibfnamefont {A.}~\bibnamefont {del Campo}},
  \bibinfo {author} {\bibfnamefont {Y.-J.}\ \bibnamefont {Han}}, \bibinfo
  {author} {\bibfnamefont {C.-F.}\ \bibnamefont {Li}},\ and\ \bibinfo {author}
  {\bibfnamefont {G.-C.}\ \bibnamefont {Guo}},\ }\bibfield  {title} {\bibinfo
  {title} {Experimental trapped-ion quantum simulation of the {K}ibble-{Z}urek
  dynamics in momentum space},\ }\href {https://doi.org/10.1038/srep33381}
  {\bibfield  {journal} {\bibinfo  {journal} {Sci. Rep.}\ }\textbf {\bibinfo
  {volume} {6}},\ \bibinfo {pages} {33381} (\bibinfo {year}
  {2016})}\BibitemShut {NoStop}%
\bibitem [{\citenamefont {Bernien}\ \emph {et~al.}(2017)\citenamefont
  {Bernien}, \citenamefont {Schwartz}, \citenamefont {Keesling}, \citenamefont
  {Levine}, \citenamefont {Omran}, \citenamefont {Pichler}, \citenamefont
  {Choi}, \citenamefont {Zibrov}, \citenamefont {Endres}, \citenamefont
  {Greiner}, \citenamefont {Vuleti\'c},\ and\ \citenamefont {Lukin}}]{Lukin17}%
  \BibitemOpen
  \bibfield  {author} {\bibinfo {author} {\bibfnamefont {H.}~\bibnamefont
  {Bernien}}, \bibinfo {author} {\bibfnamefont {S.}~\bibnamefont {Schwartz}},
  \bibinfo {author} {\bibfnamefont {A.}~\bibnamefont {Keesling}}, \bibinfo
  {author} {\bibfnamefont {H.}~\bibnamefont {Levine}}, \bibinfo {author}
  {\bibfnamefont {A.}~\bibnamefont {Omran}}, \bibinfo {author} {\bibfnamefont
  {H.}~\bibnamefont {Pichler}}, \bibinfo {author} {\bibfnamefont
  {S.}~\bibnamefont {Choi}}, \bibinfo {author} {\bibfnamefont {A.~S.}\
  \bibnamefont {Zibrov}}, \bibinfo {author} {\bibfnamefont {M.}~\bibnamefont
  {Endres}}, \bibinfo {author} {\bibfnamefont {M.}~\bibnamefont {Greiner}},
  \bibinfo {author} {\bibfnamefont {V.}~\bibnamefont {Vuleti\'c}},\ and\
  \bibinfo {author} {\bibfnamefont {M.~D.}\ \bibnamefont {Lukin}},\ }\bibfield
  {title} {\bibinfo {title} {Probing many-body dynamics on a 51-atom quantum
  simulator},\ }\href {http://dx.doi.org/10.1038/nature24622} {\bibfield
  {journal} {\bibinfo  {journal} {Nature}\ }\textbf {\bibinfo {volume} {551}},\
  \bibinfo {pages} {579} (\bibinfo {year} {2017})}\BibitemShut {NoStop}%
\bibitem [{\citenamefont {Bando}\ \emph {et~al.}(2020)\citenamefont {Bando},
  \citenamefont {Susa}, \citenamefont {Oshiyama}, \citenamefont {Shibata},
  \citenamefont {Ohzeki}, \citenamefont {G\'omez-Ruiz}, \citenamefont {Lidar},
  \citenamefont {Suzuki}, \citenamefont {del Campo},\ and\ \citenamefont
  {Nishimori}}]{Bando20}%
  \BibitemOpen
  \bibfield  {author} {\bibinfo {author} {\bibfnamefont {Y.}~\bibnamefont
  {Bando}}, \bibinfo {author} {\bibfnamefont {Y.}~\bibnamefont {Susa}},
  \bibinfo {author} {\bibfnamefont {H.}~\bibnamefont {Oshiyama}}, \bibinfo
  {author} {\bibfnamefont {N.}~\bibnamefont {Shibata}}, \bibinfo {author}
  {\bibfnamefont {M.}~\bibnamefont {Ohzeki}}, \bibinfo {author} {\bibfnamefont
  {F.~J.}\ \bibnamefont {G\'omez-Ruiz}}, \bibinfo {author} {\bibfnamefont
  {D.~A.}\ \bibnamefont {Lidar}}, \bibinfo {author} {\bibfnamefont
  {S.}~\bibnamefont {Suzuki}}, \bibinfo {author} {\bibfnamefont
  {A.}~\bibnamefont {del Campo}},\ and\ \bibinfo {author} {\bibfnamefont
  {H.}~\bibnamefont {Nishimori}},\ }\bibfield  {title} {\bibinfo {title}
  {Probing the universality of topological defect formation in a quantum
  annealer: {K}ibble-{Z}urek mechanism and beyond},\ }\href
  {https://link.aps.org/doi/10.1103/PhysRevResearch.2.033369} {\bibfield
  {journal} {\bibinfo  {journal} {Phys. Rev. Research}\ }\textbf {\bibinfo
  {volume} {2}},\ \bibinfo {pages} {033369} (\bibinfo {year}
  {2020})}\BibitemShut {NoStop}%
\bibitem [{\citenamefont {King}\ \emph {et~al.}(2022)\citenamefont {King},
  \citenamefont {Suzuki}, \citenamefont {Raymond}, \citenamefont {Zucca},
  \citenamefont {Lanting}, \citenamefont {Altomare}, \citenamefont {Berkley},
  \citenamefont {Ejtemaee}, \citenamefont {Hoskinson}, \citenamefont {Huang},
  \citenamefont {Ladizinsky}, \citenamefont {MacDonald}, \citenamefont
  {Marsden}, \citenamefont {Oh}, \citenamefont {Poulin-Lamarre}, \citenamefont
  {Reis}, \citenamefont {Rich}, \citenamefont {Sato}, \citenamefont
  {Whittaker}, \citenamefont {Yao}, \citenamefont {Harris}, \citenamefont
  {Lidar}, \citenamefont {Nishimori},\ and\ \citenamefont {Amin}}]{King2022}%
  \BibitemOpen
  \bibfield  {author} {\bibinfo {author} {\bibfnamefont {A.~D.}\ \bibnamefont
  {King}}, \bibinfo {author} {\bibfnamefont {S.}~\bibnamefont {Suzuki}},
  \bibinfo {author} {\bibfnamefont {J.}~\bibnamefont {Raymond}}, \bibinfo
  {author} {\bibfnamefont {A.}~\bibnamefont {Zucca}}, \bibinfo {author}
  {\bibfnamefont {T.}~\bibnamefont {Lanting}}, \bibinfo {author} {\bibfnamefont
  {F.}~\bibnamefont {Altomare}}, \bibinfo {author} {\bibfnamefont {A.~J.}\
  \bibnamefont {Berkley}}, \bibinfo {author} {\bibfnamefont {S.}~\bibnamefont
  {Ejtemaee}}, \bibinfo {author} {\bibfnamefont {E.}~\bibnamefont {Hoskinson}},
  \bibinfo {author} {\bibfnamefont {S.}~\bibnamefont {Huang}}, \bibinfo
  {author} {\bibfnamefont {E.}~\bibnamefont {Ladizinsky}}, \bibinfo {author}
  {\bibfnamefont {A.~J.~R.}\ \bibnamefont {MacDonald}}, \bibinfo {author}
  {\bibfnamefont {G.}~\bibnamefont {Marsden}}, \bibinfo {author} {\bibfnamefont
  {T.}~\bibnamefont {Oh}}, \bibinfo {author} {\bibfnamefont {G.}~\bibnamefont
  {Poulin-Lamarre}}, \bibinfo {author} {\bibfnamefont {M.}~\bibnamefont
  {Reis}}, \bibinfo {author} {\bibfnamefont {C.}~\bibnamefont {Rich}}, \bibinfo
  {author} {\bibfnamefont {Y.}~\bibnamefont {Sato}}, \bibinfo {author}
  {\bibfnamefont {J.~D.}\ \bibnamefont {Whittaker}}, \bibinfo {author}
  {\bibfnamefont {J.}~\bibnamefont {Yao}}, \bibinfo {author} {\bibfnamefont
  {R.}~\bibnamefont {Harris}}, \bibinfo {author} {\bibfnamefont {D.~A.}\
  \bibnamefont {Lidar}}, \bibinfo {author} {\bibfnamefont {H.}~\bibnamefont
  {Nishimori}},\ and\ \bibinfo {author} {\bibfnamefont {M.~H.}\ \bibnamefont
  {Amin}},\ }\bibfield  {title} {\bibinfo {title} {Coherent quantum annealing
  in a programmable 2,000{\thinspace}qubit ising chain},\ }\href
  {https://doi.org/10.1038/s41567-022-01741-6} {\bibfield  {journal} {\bibinfo
  {journal} {Nature Physics}\ }\textbf {\bibinfo {volume} {18}},\ \bibinfo
  {pages} {1324} (\bibinfo {year} {2022})}\BibitemShut {NoStop}%
\bibitem [{\citenamefont {Monaco}\ \emph {et~al.}(2002)\citenamefont {Monaco},
  \citenamefont {Mygind},\ and\ \citenamefont {Rivers}}]{Monaco02}%
  \BibitemOpen
  \bibfield  {author} {\bibinfo {author} {\bibfnamefont {R.}~\bibnamefont
  {Monaco}}, \bibinfo {author} {\bibfnamefont {J.}~\bibnamefont {Mygind}},\
  and\ \bibinfo {author} {\bibfnamefont {R.~J.}\ \bibnamefont {Rivers}},\
  }\bibfield  {title} {\bibinfo {title} {{Z}urek-{K}ibble domain structures:
  The dynamics of spontaneous vortex formation in annular {J}osephson tunnel
  junctions},\ }\href {https://doi.org/10.1103/PhysRevLett.89.080603}
  {\bibfield  {journal} {\bibinfo  {journal} {Phys. Rev. Lett.}\ }\textbf
  {\bibinfo {volume} {89}},\ \bibinfo {pages} {080603} (\bibinfo {year}
  {2002})}\BibitemShut {NoStop}%
\bibitem [{\citenamefont {Ejtemaee}\ and\ \citenamefont {Haljan}(2013)}]{EH13}%
  \BibitemOpen
  \bibfield  {author} {\bibinfo {author} {\bibfnamefont {S.}~\bibnamefont
  {Ejtemaee}}\ and\ \bibinfo {author} {\bibfnamefont {P.~C.}\ \bibnamefont
  {Haljan}},\ }\bibfield  {title} {\bibinfo {title} {Spontaneous nucleation and
  dynamics of kink defects in zigzag arrays of trapped ions},\ }\href
  {https://doi.org/10.1103/PhysRevA.87.051401} {\bibfield  {journal} {\bibinfo
  {journal} {Phys. Rev. A}\ }\textbf {\bibinfo {volume} {87}},\ \bibinfo
  {pages} {051401} (\bibinfo {year} {2013})}\BibitemShut {NoStop}%
\bibitem [{\citenamefont {Ulm}\ \emph {et~al.}(2013)\citenamefont {Ulm},
  \citenamefont {Ro{\ss}nagel}, \citenamefont {Jacob}, \citenamefont
  {Deg{\"u}nther}, \citenamefont {Dawkins}, \citenamefont {Poschinger},
  \citenamefont {Nigmatullin}, \citenamefont {Retzker}, \citenamefont {Plenio},
  \citenamefont {Schmidt-Kaler},\ and\ \citenamefont {Singer}}]{Ulm13}%
  \BibitemOpen
  \bibfield  {author} {\bibinfo {author} {\bibfnamefont {S.}~\bibnamefont
  {Ulm}}, \bibinfo {author} {\bibfnamefont {J.}~\bibnamefont {Ro{\ss}nagel}},
  \bibinfo {author} {\bibfnamefont {G.}~\bibnamefont {Jacob}}, \bibinfo
  {author} {\bibfnamefont {C.}~\bibnamefont {Deg{\"u}nther}}, \bibinfo {author}
  {\bibfnamefont {S.~T.}\ \bibnamefont {Dawkins}}, \bibinfo {author}
  {\bibfnamefont {U.~G.}\ \bibnamefont {Poschinger}}, \bibinfo {author}
  {\bibfnamefont {R.}~\bibnamefont {Nigmatullin}}, \bibinfo {author}
  {\bibfnamefont {A.}~\bibnamefont {Retzker}}, \bibinfo {author} {\bibfnamefont
  {M.~B.}\ \bibnamefont {Plenio}}, \bibinfo {author} {\bibfnamefont
  {F.}~\bibnamefont {Schmidt-Kaler}},\ and\ \bibinfo {author} {\bibfnamefont
  {K.}~\bibnamefont {Singer}},\ }\bibfield  {title} {\bibinfo {title}
  {Observation of the {K}ibble-{Z}urek scaling law for defect formation in ion
  crystals},\ }\href {http://dx.doi.org/10.1038/ncomms3290} {\bibfield
  {journal} {\bibinfo  {journal} {Nat. Comm.}\ }\textbf {\bibinfo {volume}
  {4}},\ \bibinfo {pages} {2290} (\bibinfo {year} {2013})}\BibitemShut
  {NoStop}%
\bibitem [{\citenamefont {Pyka}\ \emph {et~al.}(2013)\citenamefont {Pyka},
  \citenamefont {Keller}, \citenamefont {Partner}, \citenamefont {Nigmatullin},
  \citenamefont {Burgermeister}, \citenamefont {Meier}, \citenamefont
  {Kuhlmann}, \citenamefont {Retzker}, \citenamefont {Plenio}, \citenamefont
  {Zurek}, \citenamefont {del Campo},\ and\ \citenamefont
  {Mehlst\"aubler}}]{Pyka13}%
  \BibitemOpen
  \bibfield  {author} {\bibinfo {author} {\bibfnamefont {K.}~\bibnamefont
  {Pyka}}, \bibinfo {author} {\bibfnamefont {J.}~\bibnamefont {Keller}},
  \bibinfo {author} {\bibfnamefont {H.~L.}\ \bibnamefont {Partner}}, \bibinfo
  {author} {\bibfnamefont {R.}~\bibnamefont {Nigmatullin}}, \bibinfo {author}
  {\bibfnamefont {T.}~\bibnamefont {Burgermeister}}, \bibinfo {author}
  {\bibfnamefont {D.~M.}\ \bibnamefont {Meier}}, \bibinfo {author}
  {\bibfnamefont {K.}~\bibnamefont {Kuhlmann}}, \bibinfo {author}
  {\bibfnamefont {A.}~\bibnamefont {Retzker}}, \bibinfo {author} {\bibfnamefont
  {M.~B.}\ \bibnamefont {Plenio}}, \bibinfo {author} {\bibfnamefont {W.~H.}\
  \bibnamefont {Zurek}}, \bibinfo {author} {\bibfnamefont {A.}~\bibnamefont
  {del Campo}},\ and\ \bibinfo {author} {\bibfnamefont {T.~E.}\ \bibnamefont
  {Mehlst\"aubler}},\ }\bibfield  {title} {\bibinfo {title} {Topological defect
  formation and spontaneous symmetry breaking in ion coulomb crystals},\ }\href
  {http://dx.doi.org/10.1038/ncomms3291} {\bibfield  {journal} {\bibinfo
  {journal} {\href{http://dx.doi.org/10.1038/ncomms3291}{Nat. Comm.}}\ }\textbf
  {\bibinfo {volume} {4}},\ \bibinfo {pages} {2291} (\bibinfo {year}
  {2013})}\BibitemShut {NoStop}%
\bibitem [{\citenamefont {Lamporesi}\ \emph {et~al.}(2013)\citenamefont
  {Lamporesi}, \citenamefont {Donadello}, \citenamefont {Serafini},
  \citenamefont {Dalfovo},\ and\ \citenamefont {Ferrari}}]{Lamporesi13}%
  \BibitemOpen
  \bibfield  {author} {\bibinfo {author} {\bibfnamefont {G.}~\bibnamefont
  {Lamporesi}}, \bibinfo {author} {\bibfnamefont {S.}~\bibnamefont
  {Donadello}}, \bibinfo {author} {\bibfnamefont {S.}~\bibnamefont {Serafini}},
  \bibinfo {author} {\bibfnamefont {F.}~\bibnamefont {Dalfovo}},\ and\ \bibinfo
  {author} {\bibfnamefont {G.}~\bibnamefont {Ferrari}},\ }\bibfield  {title}
  {\bibinfo {title} {Spontaneous creation of {K}ibble-{Z}urek solitons in a
  bose-einstein condensate},\ }\href {http://dx.doi.org/10.1038/nphys2734}
  {\bibfield  {journal} {\bibinfo  {journal}
  {\href{http://dx.doi.org/10.1038/nphys2734}{Nature Physics}}\ }\textbf
  {\bibinfo {volume} {9}},\ \bibinfo {pages} {656} (\bibinfo {year}
  {2013})}\BibitemShut {NoStop}%
\bibitem [{\citenamefont {Casado}\ \emph {et~al.}(2001)\citenamefont {Casado},
  \citenamefont {Gonz\'alez-Vi\~nas}, \citenamefont {Mancini},\ and\
  \citenamefont {Boccaletti}}]{Casado01}%
  \BibitemOpen
  \bibfield  {author} {\bibinfo {author} {\bibfnamefont {S.}~\bibnamefont
  {Casado}}, \bibinfo {author} {\bibfnamefont {W.}~\bibnamefont
  {Gonz\'alez-Vi\~nas}}, \bibinfo {author} {\bibfnamefont {H.}~\bibnamefont
  {Mancini}},\ and\ \bibinfo {author} {\bibfnamefont {S.}~\bibnamefont
  {Boccaletti}},\ }\bibfield  {title} {\bibinfo {title} {Topological defects
  after a quench in a {B}\'enard-{M}arangoni convection system},\ }\href
  {https://doi.org/10.1103/PhysRevE.63.057301} {\bibfield  {journal} {\bibinfo
  {journal} {Phys. Rev. E}\ }\textbf {\bibinfo {volume} {63}},\ \bibinfo
  {pages} {057301} (\bibinfo {year} {2001})}\BibitemShut {NoStop}%
\bibitem [{\citenamefont {Casado}\ \emph {et~al.}(2006)\citenamefont {Casado},
  \citenamefont {Gonz\'alez-Vi\~nas},\ and\ \citenamefont
  {Mancini}}]{Casado06}%
  \BibitemOpen
  \bibfield  {author} {\bibinfo {author} {\bibfnamefont {S.}~\bibnamefont
  {Casado}}, \bibinfo {author} {\bibfnamefont {W.}~\bibnamefont
  {Gonz\'alez-Vi\~nas}},\ and\ \bibinfo {author} {\bibfnamefont
  {H.}~\bibnamefont {Mancini}},\ }\bibfield  {title} {\bibinfo {title} {Testing
  the {K}ibble-{Z}urek mechanism in {R}ayleigh-{B}\'enard convection},\ }\href
  {https://doi.org/10.1103/PhysRevE.74.047101} {\bibfield  {journal} {\bibinfo
  {journal} {Phys. Rev. E}\ }\textbf {\bibinfo {volume} {74}},\ \bibinfo
  {pages} {047101} (\bibinfo {year} {2006})}\BibitemShut {NoStop}%
\bibitem [{\citenamefont {Chae}\ \emph {et~al.}(2012)\citenamefont {Chae},
  \citenamefont {Lee}, \citenamefont {Horibe}, \citenamefont {Tanimura},
  \citenamefont {Mori}, \citenamefont {Gao}, \citenamefont {Carr},\ and\
  \citenamefont {Cheong}}]{Chae12}%
  \BibitemOpen
  \bibfield  {author} {\bibinfo {author} {\bibfnamefont {S.~C.}\ \bibnamefont
  {Chae}}, \bibinfo {author} {\bibfnamefont {N.}~\bibnamefont {Lee}}, \bibinfo
  {author} {\bibfnamefont {Y.}~\bibnamefont {Horibe}}, \bibinfo {author}
  {\bibfnamefont {M.}~\bibnamefont {Tanimura}}, \bibinfo {author}
  {\bibfnamefont {S.}~\bibnamefont {Mori}}, \bibinfo {author} {\bibfnamefont
  {B.}~\bibnamefont {Gao}}, \bibinfo {author} {\bibfnamefont {S.}~\bibnamefont
  {Carr}},\ and\ \bibinfo {author} {\bibfnamefont {S.-W.}\ \bibnamefont
  {Cheong}},\ }\bibfield  {title} {\bibinfo {title} {Direct observation of the
  proliferation of ferroelectric loop domains and vortex-antivortex pairs},\
  }\href {https://doi.org/10.1103/PhysRevLett.108.167603} {\bibfield  {journal}
  {\bibinfo  {journal} {Phys. Rev. Lett.}\ }\textbf {\bibinfo {volume} {108}},\
  \bibinfo {pages} {167603} (\bibinfo {year} {2012})}\BibitemShut {NoStop}%
\bibitem [{\citenamefont {Lin}\ \emph {et~al.}(2014)\citenamefont {Lin},
  \citenamefont {Wang}, \citenamefont {Kamiya}, \citenamefont {Chern},
  \citenamefont {Fan}, \citenamefont {Fan}, \citenamefont {Casas},
  \citenamefont {Liu}, \citenamefont {Kiryukhin}, \citenamefont {Zurek},
  \citenamefont {Batista},\ and\ \citenamefont {Cheong}}]{Lin14}%
  \BibitemOpen
  \bibfield  {author} {\bibinfo {author} {\bibfnamefont {S.-Z.}\ \bibnamefont
  {Lin}}, \bibinfo {author} {\bibfnamefont {X.}~\bibnamefont {Wang}}, \bibinfo
  {author} {\bibfnamefont {Y.}~\bibnamefont {Kamiya}}, \bibinfo {author}
  {\bibfnamefont {G.-W.}\ \bibnamefont {Chern}}, \bibinfo {author}
  {\bibfnamefont {F.}~\bibnamefont {Fan}}, \bibinfo {author} {\bibfnamefont
  {D.}~\bibnamefont {Fan}}, \bibinfo {author} {\bibfnamefont {B.}~\bibnamefont
  {Casas}}, \bibinfo {author} {\bibfnamefont {Y.}~\bibnamefont {Liu}}, \bibinfo
  {author} {\bibfnamefont {V.}~\bibnamefont {Kiryukhin}}, \bibinfo {author}
  {\bibfnamefont {W.~H.}\ \bibnamefont {Zurek}}, \bibinfo {author}
  {\bibfnamefont {C.~D.}\ \bibnamefont {Batista}},\ and\ \bibinfo {author}
  {\bibfnamefont {S.-W.}\ \bibnamefont {Cheong}},\ }\bibfield  {title}
  {\bibinfo {title} {Topological defects as relics of emergent continuous
  symmetry and {H}iggs condensation of disorder in ferroelectrics},\ }\href
  {https://doi.org/10.1038/nphys3142} {\bibfield  {journal} {\bibinfo
  {journal} {Nature Physics}\ }\textbf {\bibinfo {volume} {10}},\ \bibinfo
  {pages} {970} (\bibinfo {year} {2014})}\BibitemShut {NoStop}%
\bibitem [{\citenamefont {Navon}\ \emph {et~al.}(2015)\citenamefont {Navon},
  \citenamefont {Gaunt}, \citenamefont {Smith},\ and\ \citenamefont
  {Hadzibabic}}]{Navon15}%
  \BibitemOpen
  \bibfield  {author} {\bibinfo {author} {\bibfnamefont {N.}~\bibnamefont
  {Navon}}, \bibinfo {author} {\bibfnamefont {A.~L.}\ \bibnamefont {Gaunt}},
  \bibinfo {author} {\bibfnamefont {R.~P.}\ \bibnamefont {Smith}},\ and\
  \bibinfo {author} {\bibfnamefont {Z.}~\bibnamefont {Hadzibabic}},\ }\bibfield
   {title} {\bibinfo {title} {Critical dynamics of spontaneous symmetry
  breaking in a homogeneous {B}ose gas},\ }\href
  {https://doi.org/10.1126/science.1258676} {\bibfield  {journal} {\bibinfo
  {journal} {Science}\ }\textbf {\bibinfo {volume} {347}},\ \bibinfo {pages}
  {167} (\bibinfo {year} {2015})}\BibitemShut {NoStop}%
\bibitem [{\citenamefont {Chomaz}\ \emph {et~al.}(2015)\citenamefont {Chomaz},
  \citenamefont {Corman}, \citenamefont {Bienaim{\'e}}, \citenamefont
  {Desbuquois}, \citenamefont {Weitenberg}, \citenamefont {Nascimb{\`e}ne},
  \citenamefont {Beugnon},\ and\ \citenamefont {Dalibard}}]{Chomaz15}%
  \BibitemOpen
  \bibfield  {author} {\bibinfo {author} {\bibfnamefont {L.}~\bibnamefont
  {Chomaz}}, \bibinfo {author} {\bibfnamefont {L.}~\bibnamefont {Corman}},
  \bibinfo {author} {\bibfnamefont {T.}~\bibnamefont {Bienaim{\'e}}}, \bibinfo
  {author} {\bibfnamefont {R.}~\bibnamefont {Desbuquois}}, \bibinfo {author}
  {\bibfnamefont {C.}~\bibnamefont {Weitenberg}}, \bibinfo {author}
  {\bibfnamefont {S.}~\bibnamefont {Nascimb{\`e}ne}}, \bibinfo {author}
  {\bibfnamefont {J.}~\bibnamefont {Beugnon}},\ and\ \bibinfo {author}
  {\bibfnamefont {J.}~\bibnamefont {Dalibard}},\ }\bibfield  {title} {\bibinfo
  {title} {Emergence of coherence via transverse condensation in a uniform
  quasi-two-dimensional {B}ose gas},\ }\href
  {https://doi.org/10.1038/ncomms7162} {\bibfield  {journal} {\bibinfo
  {journal} {Nat. Comm.}\ }\textbf {\bibinfo {volume} {6}},\ \bibinfo {pages}
  {6162} (\bibinfo {year} {2015})}\BibitemShut {NoStop}%
\bibitem [{\citenamefont {Meier}\ \emph {et~al.}(2017)\citenamefont {Meier},
  \citenamefont {Lilienblum}, \citenamefont {Griffin}, \citenamefont {Conder},
  \citenamefont {Pomjakushina}, \citenamefont {Yan}, \citenamefont {Bourret},
  \citenamefont {Meier}, \citenamefont {Lichtenberg}, \citenamefont {Salje},
  \citenamefont {Spaldin}, \citenamefont {Fiebig},\ and\ \citenamefont
  {Cano}}]{Meier17}%
  \BibitemOpen
  \bibfield  {author} {\bibinfo {author} {\bibfnamefont {Q.~N.}\ \bibnamefont
  {Meier}}, \bibinfo {author} {\bibfnamefont {M.}~\bibnamefont {Lilienblum}},
  \bibinfo {author} {\bibfnamefont {S.~M.}\ \bibnamefont {Griffin}}, \bibinfo
  {author} {\bibfnamefont {K.}~\bibnamefont {Conder}}, \bibinfo {author}
  {\bibfnamefont {E.}~\bibnamefont {Pomjakushina}}, \bibinfo {author}
  {\bibfnamefont {Z.}~\bibnamefont {Yan}}, \bibinfo {author} {\bibfnamefont
  {E.}~\bibnamefont {Bourret}}, \bibinfo {author} {\bibfnamefont
  {D.}~\bibnamefont {Meier}}, \bibinfo {author} {\bibfnamefont
  {F.}~\bibnamefont {Lichtenberg}}, \bibinfo {author} {\bibfnamefont
  {E.~K.~H.}\ \bibnamefont {Salje}}, \bibinfo {author} {\bibfnamefont {N.~A.}\
  \bibnamefont {Spaldin}}, \bibinfo {author} {\bibfnamefont {M.}~\bibnamefont
  {Fiebig}},\ and\ \bibinfo {author} {\bibfnamefont {A.}~\bibnamefont {Cano}},\
  }\bibfield  {title} {\bibinfo {title} {Global formation of topological
  defects in the multiferroic hexagonal manganites},\ }\href
  {https://doi.org/10.1103/PhysRevX.7.041014} {\bibfield  {journal} {\bibinfo
  {journal} {Phys. Rev. X}\ }\textbf {\bibinfo {volume} {7}},\ \bibinfo {pages}
  {041014} (\bibinfo {year} {2017})}\BibitemShut {NoStop}%
\bibitem [{\citenamefont {Ko}\ \emph {et~al.}(2019)\citenamefont {Ko},
  \citenamefont {Park},\ and\ \citenamefont {Shin}}]{Shin19}%
  \BibitemOpen
  \bibfield  {author} {\bibinfo {author} {\bibfnamefont {B.}~\bibnamefont
  {Ko}}, \bibinfo {author} {\bibfnamefont {J.~W.}\ \bibnamefont {Park}},\ and\
  \bibinfo {author} {\bibfnamefont {Y.}~\bibnamefont {Shin}},\ }\bibfield
  {title} {\bibinfo {title} {{K}ibble-{Z}urek universality in a strongly
  interacting fermi superfluid},\ }\href
  {https://doi.org/10.1038/s41567-019-0650-1} {\bibfield  {journal} {\bibinfo
  {journal} {Nature Physics}\ }\textbf {\bibinfo {volume} {15}},\ \bibinfo
  {pages} {1227} (\bibinfo {year} {2019})}\BibitemShut {NoStop}%
\bibitem [{\citenamefont {Goo}\ \emph {et~al.}(2021)\citenamefont {Goo},
  \citenamefont {Lim},\ and\ \citenamefont {Shin}}]{Goo21}%
  \BibitemOpen
  \bibfield  {author} {\bibinfo {author} {\bibfnamefont {J.}~\bibnamefont
  {Goo}}, \bibinfo {author} {\bibfnamefont {Y.}~\bibnamefont {Lim}},\ and\
  \bibinfo {author} {\bibfnamefont {Y.}~\bibnamefont {Shin}},\ }\bibfield
  {title} {\bibinfo {title} {Defect saturation in a rapidly quenched bose
  gas},\ }\href {https://doi.org/10.1103/PhysRevLett.127.115701} {\bibfield
  {journal} {\bibinfo  {journal} {Phys. Rev. Lett.}\ }\textbf {\bibinfo
  {volume} {127}},\ \bibinfo {pages} {115701} (\bibinfo {year}
  {2021})}\BibitemShut {NoStop}%
\bibitem [{\citenamefont {Kim}\ \emph {et~al.}(2022)\citenamefont {Kim},
  \citenamefont {Rabga}, \citenamefont {Lee}, \citenamefont {Goo},
  \citenamefont {Bae},\ and\ \citenamefont {Shin}}]{KimShin22}%
  \BibitemOpen
  \bibfield  {author} {\bibinfo {author} {\bibfnamefont {M.}~\bibnamefont
  {Kim}}, \bibinfo {author} {\bibfnamefont {T.}~\bibnamefont {Rabga}}, \bibinfo
  {author} {\bibfnamefont {Y.}~\bibnamefont {Lee}}, \bibinfo {author}
  {\bibfnamefont {J.}~\bibnamefont {Goo}}, \bibinfo {author} {\bibfnamefont
  {D.}~\bibnamefont {Bae}},\ and\ \bibinfo {author} {\bibfnamefont
  {Y.}~\bibnamefont {Shin}},\ }\bibfield  {title} {\bibinfo {title}
  {Suppression of spontaneous defect formation in inhomogeneous bose gases},\
  }\href {https://doi.org/10.1103/PhysRevA.106.L061301} {\bibfield  {journal}
  {\bibinfo  {journal} {Phys. Rev. A}\ }\textbf {\bibinfo {volume} {106}},\
  \bibinfo {pages} {L061301} (\bibinfo {year} {2022})}\BibitemShut {NoStop}%
\bibitem [{\citenamefont {Du}\ \emph {et~al.}(2023)\citenamefont {Du},
  \citenamefont {Fang}, \citenamefont {Won}, \citenamefont {De}, \citenamefont
  {Huang}, \citenamefont {Xu}, \citenamefont {You}, \citenamefont
  {G{\'o}mez-Ruiz}, \citenamefont {del Campo},\ and\ \citenamefont
  {Cheong}}]{Du23}%
  \BibitemOpen
  \bibfield  {author} {\bibinfo {author} {\bibfnamefont {K.}~\bibnamefont
  {Du}}, \bibinfo {author} {\bibfnamefont {X.}~\bibnamefont {Fang}}, \bibinfo
  {author} {\bibfnamefont {C.}~\bibnamefont {Won}}, \bibinfo {author}
  {\bibfnamefont {C.}~\bibnamefont {De}}, \bibinfo {author} {\bibfnamefont
  {F.-T.}\ \bibnamefont {Huang}}, \bibinfo {author} {\bibfnamefont
  {W.}~\bibnamefont {Xu}}, \bibinfo {author} {\bibfnamefont {H.}~\bibnamefont
  {You}}, \bibinfo {author} {\bibfnamefont {F.~J.}\ \bibnamefont
  {G{\'o}mez-Ruiz}}, \bibinfo {author} {\bibfnamefont {A.}~\bibnamefont {del
  Campo}},\ and\ \bibinfo {author} {\bibfnamefont {S.-W.}\ \bibnamefont
  {Cheong}},\ }\bibfield  {title} {\bibinfo {title} {Kibble--{Z}urek mechanism
  of ising domains},\ }\bibfield  {journal} {\bibinfo  {journal} {Nature
  Physics}\ }\href {https://doi.org/10.1038/s41567-023-02112-5}
  {10.1038/s41567-023-02112-5} (\bibinfo {year} {2023})\BibitemShut {NoStop}%
\bibitem [{\citenamefont {Jeli{\'{c}}}\ and\ \citenamefont
  {Cugliandolo}(2011)}]{Jeli11}%
  \BibitemOpen
  \bibfield  {author} {\bibinfo {author} {\bibfnamefont {A.}~\bibnamefont
  {Jeli{\'{c}}}}\ and\ \bibinfo {author} {\bibfnamefont {L.~F.}\ \bibnamefont
  {Cugliandolo}},\ }\bibfield  {title} {\bibinfo {title} {Quench dynamics of
  the 2dxymodel},\ }\href {https://doi.org/10.1088/1742-5468/2011/02/p02032}
  {\bibfield  {journal} {\bibinfo  {journal} {Journal of Statistical Mechanics:
  Theory and Experiment}\ }\textbf {\bibinfo {volume} {2011}},\ \bibinfo
  {pages} {P02032} (\bibinfo {year} {2011})}\BibitemShut {NoStop}%
\bibitem [{\citenamefont {Griffin}\ \emph {et~al.}(2012)\citenamefont
  {Griffin}, \citenamefont {Lilienblum}, \citenamefont {Delaney}, \citenamefont
  {Kumagai}, \citenamefont {Fiebig},\ and\ \citenamefont
  {Spaldin}}]{Griffin12}%
  \BibitemOpen
  \bibfield  {author} {\bibinfo {author} {\bibfnamefont {S.~M.}\ \bibnamefont
  {Griffin}}, \bibinfo {author} {\bibfnamefont {M.}~\bibnamefont {Lilienblum}},
  \bibinfo {author} {\bibfnamefont {K.~T.}\ \bibnamefont {Delaney}}, \bibinfo
  {author} {\bibfnamefont {Y.}~\bibnamefont {Kumagai}}, \bibinfo {author}
  {\bibfnamefont {M.}~\bibnamefont {Fiebig}},\ and\ \bibinfo {author}
  {\bibfnamefont {N.~A.}\ \bibnamefont {Spaldin}},\ }\bibfield  {title}
  {\bibinfo {title} {Scaling behavior and beyond equilibrium in the hexagonal
  manganites},\ }\href {https://doi.org/10.1103/PhysRevX.2.041022} {\bibfield
  {journal} {\bibinfo  {journal} {Phys. Rev. X}\ }\textbf {\bibinfo {volume}
  {2}},\ \bibinfo {pages} {041022} (\bibinfo {year} {2012})}\BibitemShut
  {NoStop}%
\bibitem [{\citenamefont {Deutschl{\"a}nder}\ \emph {et~al.}(2015)\citenamefont
  {Deutschl{\"a}nder}, \citenamefont {Dillmann}, \citenamefont {Maret},\ and\
  \citenamefont {Keim}}]{Keim15}%
  \BibitemOpen
  \bibfield  {author} {\bibinfo {author} {\bibfnamefont {S.}~\bibnamefont
  {Deutschl{\"a}nder}}, \bibinfo {author} {\bibfnamefont {P.}~\bibnamefont
  {Dillmann}}, \bibinfo {author} {\bibfnamefont {G.}~\bibnamefont {Maret}},\
  and\ \bibinfo {author} {\bibfnamefont {P.}~\bibnamefont {Keim}},\ }\bibfield
  {title} {\bibinfo {title} {{K}ibble-{Z}urek mechanism in colloidal
  monolayers},\ }\href {https://doi.org/10.1073/pnas.1500763112} {\bibfield
  {journal} {\bibinfo  {journal} {Proc. Nat. Acad. Sciences}\ }\textbf
  {\bibinfo {volume} {112}},\ \bibinfo {pages} {6925} (\bibinfo {year}
  {2015})}\BibitemShut {NoStop}%
\bibitem [{\citenamefont {Chesler}\ \emph {et~al.}(2015)\citenamefont
  {Chesler}, \citenamefont {Garc\'{\i}a-Garc\'{\i}a},\ and\ \citenamefont
  {Liu}}]{Chesler15}%
  \BibitemOpen
  \bibfield  {author} {\bibinfo {author} {\bibfnamefont {P.~M.}\ \bibnamefont
  {Chesler}}, \bibinfo {author} {\bibfnamefont {A.~M.}\ \bibnamefont
  {Garc\'{\i}a-Garc\'{\i}a}},\ and\ \bibinfo {author} {\bibfnamefont
  {H.}~\bibnamefont {Liu}},\ }\bibfield  {title} {\bibinfo {title} {Defect
  formation beyond {K}ibble-{Z}urek mechanism and holography},\ }\href
  {https://doi.org/10.1103/PhysRevX.5.021015} {\bibfield  {journal} {\bibinfo
  {journal} {Phys. Rev. X}\ }\textbf {\bibinfo {volume} {5}},\ \bibinfo {pages}
  {021015} (\bibinfo {year} {2015})}\BibitemShut {NoStop}%
\bibitem [{\citenamefont {Biroli}\ \emph {et~al.}(2010)\citenamefont {Biroli},
  \citenamefont {Cugliandolo},\ and\ \citenamefont {Sicilia}}]{Biroli10}%
  \BibitemOpen
  \bibfield  {author} {\bibinfo {author} {\bibfnamefont {G.}~\bibnamefont
  {Biroli}}, \bibinfo {author} {\bibfnamefont {L.~F.}\ \bibnamefont
  {Cugliandolo}},\ and\ \bibinfo {author} {\bibfnamefont {A.}~\bibnamefont
  {Sicilia}},\ }\bibfield  {title} {\bibinfo {title} {Kibble-{Z}urek mechanism
  and infinitely slow annealing through critical points},\ }\href
  {https://doi.org/10.1103/PhysRevE.81.050101} {\bibfield  {journal} {\bibinfo
  {journal} {Phys. Rev. E}\ }\textbf {\bibinfo {volume} {81}},\ \bibinfo
  {pages} {050101} (\bibinfo {year} {2010})}\BibitemShut {NoStop}%
\bibitem [{\citenamefont {Lib\'al}\ \emph {et~al.}(2020)\citenamefont
  {Lib\'al}, \citenamefont {del Campo}, \citenamefont {Nisoli}, \citenamefont
  {Reichhardt},\ and\ \citenamefont {Reichhardt}}]{Libal20}%
  \BibitemOpen
  \bibfield  {author} {\bibinfo {author} {\bibfnamefont {A.}~\bibnamefont
  {Lib\'al}}, \bibinfo {author} {\bibfnamefont {A.}~\bibnamefont {del Campo}},
  \bibinfo {author} {\bibfnamefont {C.}~\bibnamefont {Nisoli}}, \bibinfo
  {author} {\bibfnamefont {C.}~\bibnamefont {Reichhardt}},\ and\ \bibinfo
  {author} {\bibfnamefont {C.~J.~O.}\ \bibnamefont {Reichhardt}},\ }\bibfield
  {title} {\bibinfo {title} {Quenched dynamics of artificial colloidal spin
  ice},\ }\href {https://doi.org/10.1103/PhysRevResearch.2.033433} {\bibfield
  {journal} {\bibinfo  {journal} {Phys. Rev. Res.}\ }\textbf {\bibinfo {volume}
  {2}},\ \bibinfo {pages} {033433} (\bibinfo {year} {2020})}\BibitemShut
  {NoStop}%
\bibitem [{\citenamefont {Snyder}\ \emph {et~al.}(2004)\citenamefont {Snyder},
  \citenamefont {Ueland}, \citenamefont {Slusky}, \citenamefont {Karunadasa},
  \citenamefont {Cava},\ and\ \citenamefont {Schiffer}}]{Snyder04}%
  \BibitemOpen
  \bibfield  {author} {\bibinfo {author} {\bibfnamefont {J.}~\bibnamefont
  {Snyder}}, \bibinfo {author} {\bibfnamefont {B.~G.}\ \bibnamefont {Ueland}},
  \bibinfo {author} {\bibfnamefont {J.~S.}\ \bibnamefont {Slusky}}, \bibinfo
  {author} {\bibfnamefont {H.}~\bibnamefont {Karunadasa}}, \bibinfo {author}
  {\bibfnamefont {R.~J.}\ \bibnamefont {Cava}},\ and\ \bibinfo {author}
  {\bibfnamefont {P.}~\bibnamefont {Schiffer}},\ }\bibfield  {title} {\bibinfo
  {title} {Low-temperature spin freezing in the {D}y$_2${T}i$_2${O}$_7$ spin
  ice},\ }\href {https://doi.org/10.1103/PhysRevB.69.064414} {\bibfield
  {journal} {\bibinfo  {journal} {Phys. Rev. B}\ }\textbf {\bibinfo {volume}
  {69}},\ \bibinfo {pages} {064414} (\bibinfo {year} {2004})}\BibitemShut
  {NoStop}%
\bibitem [{\citenamefont {Harris}\ \emph {et~al.}(1997)\citenamefont {Harris},
  \citenamefont {Bramwell}, \citenamefont {McMorrow}, \citenamefont {Zeiske},\
  and\ \citenamefont {Godfrey}}]{Harris97}%
  \BibitemOpen
  \bibfield  {author} {\bibinfo {author} {\bibfnamefont {M.~J.}\ \bibnamefont
  {Harris}}, \bibinfo {author} {\bibfnamefont {S.~T.}\ \bibnamefont
  {Bramwell}}, \bibinfo {author} {\bibfnamefont {D.~F.}\ \bibnamefont
  {McMorrow}}, \bibinfo {author} {\bibfnamefont {T.}~\bibnamefont {Zeiske}},\
  and\ \bibinfo {author} {\bibfnamefont {K.~W.}\ \bibnamefont {Godfrey}},\
  }\bibfield  {title} {\bibinfo {title} {Geometrical frustration in the
  ferromagnetic pyrochlore {H}o$_2${T}i$_2${O}$_7$},\ }\href
  {https://doi.org/10.1103/PhysRevLett.79.2554} {\bibfield  {journal} {\bibinfo
   {journal} {Phys. Rev. Lett.}\ }\textbf {\bibinfo {volume} {79}},\ \bibinfo
  {pages} {2554} (\bibinfo {year} {1997})}\BibitemShut {NoStop}%
\bibitem [{\citenamefont {Bramwell}\ and\ \citenamefont
  {Gingras}(2001)}]{Bramwell01}%
  \BibitemOpen
  \bibfield  {author} {\bibinfo {author} {\bibfnamefont {S.~T.}\ \bibnamefont
  {Bramwell}}\ and\ \bibinfo {author} {\bibfnamefont {M.~J.~P.}\ \bibnamefont
  {Gingras}},\ }\bibfield  {title} {\bibinfo {title} {Spin ice state in
  frustrated magnetic pyrochlore materials},\ }\href
  {https://doi.org/10.1126/science.1064761} {\bibfield  {journal} {\bibinfo
  {journal} {Science}\ }\textbf {\bibinfo {volume} {294}},\ \bibinfo {pages}
  {1495} (\bibinfo {year} {2001})}\BibitemShut {NoStop}%
\bibitem [{\citenamefont {den Hertog}\ and\ \citenamefont
  {Gingras}(2000)}]{Hertog00}%
  \BibitemOpen
  \bibfield  {author} {\bibinfo {author} {\bibfnamefont {B.~C.}\ \bibnamefont
  {den Hertog}}\ and\ \bibinfo {author} {\bibfnamefont {M.~J.~P.}\ \bibnamefont
  {Gingras}},\ }\bibfield  {title} {\bibinfo {title} {Dipolar interactions and
  origin of spin ice in ising pyrochlore magnets},\ }\href
  {https://doi.org/10.1103/PhysRevLett.84.3430} {\bibfield  {journal} {\bibinfo
   {journal} {Phys. Rev. Lett.}\ }\textbf {\bibinfo {volume} {84}},\ \bibinfo
  {pages} {3430} (\bibinfo {year} {2000})}\BibitemShut {NoStop}%
\bibitem [{\citenamefont {Castelnovo}\ \emph {et~al.}(2008)\citenamefont
  {Castelnovo}, \citenamefont {Moessner},\ and\ \citenamefont
  {Sondhi}}]{Castelnovo2008}%
  \BibitemOpen
  \bibfield  {author} {\bibinfo {author} {\bibfnamefont {C.}~\bibnamefont
  {Castelnovo}}, \bibinfo {author} {\bibfnamefont {R.}~\bibnamefont
  {Moessner}},\ and\ \bibinfo {author} {\bibfnamefont {S.~L.}\ \bibnamefont
  {Sondhi}},\ }\bibfield  {title} {\bibinfo {title} {Magnetic monopoles in spin
  ice},\ }\href {https://doi.org/10.1038/nature06433} {\bibfield  {journal}
  {\bibinfo  {journal} {Nature}\ }\textbf {\bibinfo {volume} {451}},\ \bibinfo
  {pages} {42} (\bibinfo {year} {2008})}\BibitemShut {NoStop}%
\bibitem [{\citenamefont {Pauling}(1935)}]{Pauling35}%
  \BibitemOpen
  \bibfield  {author} {\bibinfo {author} {\bibfnamefont {L.}~\bibnamefont
  {Pauling}},\ }\bibfield  {title} {\bibinfo {title} {The structure and entropy
  of ice and of other crystals with some randomness of atomic arrangement},\
  }\href {https://doi.org/10.1021/ja01315a102} {\bibfield  {journal} {\bibinfo
  {journal} {Journal of the American Chemical Society}\ }\textbf {\bibinfo
  {volume} {57}},\ \bibinfo {pages} {2680} (\bibinfo {year}
  {1935})}\BibitemShut {NoStop}%
\bibitem [{\citenamefont {Ramirez}\ \emph {et~al.}(1999)\citenamefont
  {Ramirez}, \citenamefont {Hayashi}, \citenamefont {Cava}, \citenamefont
  {Siddharthan},\ and\ \citenamefont {Shastry}}]{Ramirez99}%
  \BibitemOpen
  \bibfield  {author} {\bibinfo {author} {\bibfnamefont {A.~P.}\ \bibnamefont
  {Ramirez}}, \bibinfo {author} {\bibfnamefont {A.}~\bibnamefont {Hayashi}},
  \bibinfo {author} {\bibfnamefont {R.~J.}\ \bibnamefont {Cava}}, \bibinfo
  {author} {\bibfnamefont {R.}~\bibnamefont {Siddharthan}},\ and\ \bibinfo
  {author} {\bibfnamefont {B.~S.}\ \bibnamefont {Shastry}},\ }\bibfield
  {title} {\bibinfo {title} {Zero-point entropy in `spin ice'},\ }\href
  {https://doi.org/10.1038/20619} {\bibfield  {journal} {\bibinfo  {journal}
  {Nature}\ }\textbf {\bibinfo {volume} {399}},\ \bibinfo {pages} {333}
  (\bibinfo {year} {1999})}\BibitemShut {NoStop}%
\bibitem [{\citenamefont {Jaubert}\ and\ \citenamefont
  {Holdsworth}(2009)}]{Jaubert_2009}%
  \BibitemOpen
  \bibfield  {author} {\bibinfo {author} {\bibfnamefont {L.~D.~C.}\
  \bibnamefont {Jaubert}}\ and\ \bibinfo {author} {\bibfnamefont {P.~C.~W.}\
  \bibnamefont {Holdsworth}},\ }\bibfield  {title} {\bibinfo {title} {Signature
  of magnetic monopole and dirac string dynamics in spin ice},\ }\href
  {https://doi.org/10.1038/nphys1227} {\bibfield  {journal} {\bibinfo
  {journal} {Nature Physics}\ }\textbf {\bibinfo {volume} {5}},\ \bibinfo
  {pages} {258} (\bibinfo {year} {2009})}\BibitemShut {NoStop}%
\bibitem [{\citenamefont {Isakov}\ \emph {et~al.}(2004)\citenamefont {Isakov},
  \citenamefont {Gregor}, \citenamefont {Moessner},\ and\ \citenamefont
  {Sondhi}}]{isakov04}%
  \BibitemOpen
  \bibfield  {author} {\bibinfo {author} {\bibfnamefont {S.~V.}\ \bibnamefont
  {Isakov}}, \bibinfo {author} {\bibfnamefont {K.}~\bibnamefont {Gregor}},
  \bibinfo {author} {\bibfnamefont {R.}~\bibnamefont {Moessner}},\ and\
  \bibinfo {author} {\bibfnamefont {S.~L.}\ \bibnamefont {Sondhi}},\ }\bibfield
   {title} {\bibinfo {title} {Dipolar spin correlations in classical pyrochlore
  magnets},\ }\href {https://doi.org/10.1103/PhysRevLett.93.167204} {\bibfield
  {journal} {\bibinfo  {journal} {Phys. Rev. Lett.}\ }\textbf {\bibinfo
  {volume} {93}},\ \bibinfo {pages} {167204} (\bibinfo {year}
  {2004})}\BibitemShut {NoStop}%
\bibitem [{\citenamefont {Henley}(2005)}]{henley05}%
  \BibitemOpen
  \bibfield  {author} {\bibinfo {author} {\bibfnamefont {C.~L.}\ \bibnamefont
  {Henley}},\ }\bibfield  {title} {\bibinfo {title} {Power-law spin
  correlations in pyrochlore antiferromagnets},\ }\href
  {https://doi.org/10.1103/PhysRevB.71.014424} {\bibfield  {journal} {\bibinfo
  {journal} {Phys. Rev. B}\ }\textbf {\bibinfo {volume} {71}},\ \bibinfo
  {pages} {014424} (\bibinfo {year} {2005})}\BibitemShut {NoStop}%
\bibitem [{\citenamefont {Ryzhkin}(2005)}]{Ryzhkin05}%
  \BibitemOpen
  \bibfield  {author} {\bibinfo {author} {\bibfnamefont {I.~A.}\ \bibnamefont
  {Ryzhkin}},\ }\bibfield  {title} {\bibinfo {title} {Magnetic relaxation in
  rare-earth oxide pyrochlores},\ }\href {https://doi.org/10.1134/1.2103216}
  {\bibfield  {journal} {\bibinfo  {journal} {Journal of Experimental and
  Theoretical Physics}\ }\textbf {\bibinfo {volume} {101}},\ \bibinfo {pages}
  {481} (\bibinfo {year} {2005})}\BibitemShut {NoStop}%
\bibitem [{\citenamefont {Glauber}(1963)}]{Glauber1963}%
  \BibitemOpen
  \bibfield  {author} {\bibinfo {author} {\bibfnamefont {R.~J.}\ \bibnamefont
  {Glauber}},\ }\bibfield  {title} {\bibinfo {title} {Time‐dependent
  statistics of the ising model},\ }\href {https://doi.org/10.1063/1.1703954}
  {\bibfield  {journal} {\bibinfo  {journal} {Journal of Mathematical Physics}\
  }\textbf {\bibinfo {volume} {4}},\ \bibinfo {pages} {294} (\bibinfo {year}
  {1963})}\BibitemShut {NoStop}%
\bibitem [{\citenamefont {Krapivsky}\ \emph {et~al.}(2010)\citenamefont
  {Krapivsky}, \citenamefont {Redner},\ and\ \citenamefont
  {Ben-Naim}}]{NaimBook}%
  \BibitemOpen
  \bibfield  {author} {\bibinfo {author} {\bibfnamefont {P.~L.}\ \bibnamefont
  {Krapivsky}}, \bibinfo {author} {\bibfnamefont {S.}~\bibnamefont {Redner}},\
  and\ \bibinfo {author} {\bibfnamefont {E.}~\bibnamefont {Ben-Naim}},\
  }\href@noop {} {\emph {\bibinfo {title} {A Kinetic View of Statistical
  Physics}}}\ (\bibinfo  {publisher} {Cambridge University Press},\ \bibinfo
  {year} {2010})\BibitemShut {NoStop}%
\bibitem [{\citenamefont {Jeong}\ \emph {et~al.}(2020)\citenamefont {Jeong},
  \citenamefont {Kim},\ and\ \citenamefont {Lee}}]{Lee2020}%
  \BibitemOpen
  \bibfield  {author} {\bibinfo {author} {\bibfnamefont {K.}~\bibnamefont
  {Jeong}}, \bibinfo {author} {\bibfnamefont {B.}~\bibnamefont {Kim}},\ and\
  \bibinfo {author} {\bibfnamefont {S.~J.}\ \bibnamefont {Lee}},\ }\bibfield
  {title} {\bibinfo {title} {Nonequilibrium kinetics of excess defect
  generation and dynamic scaling in the ising spin chain under slow cooling},\
  }\href {https://doi.org/10.1103/PhysRevE.102.012114} {\bibfield  {journal}
  {\bibinfo  {journal} {Phys. Rev. E}\ }\textbf {\bibinfo {volume} {102}},\
  \bibinfo {pages} {012114} (\bibinfo {year} {2020})}\BibitemShut {NoStop}%
\bibitem [{\citenamefont {Krapivsky}(2010)}]{Krapivsky2010}%
  \BibitemOpen
  \bibfield  {author} {\bibinfo {author} {\bibfnamefont {P.~L.}\ \bibnamefont
  {Krapivsky}},\ }\bibfield  {title} {\bibinfo {title} {Slow cooling of an
  ising ferromagnet},\ }\href
  {https://doi.org/10.1088/1742-5468/2010/02/p02014} {\bibfield  {journal}
  {\bibinfo  {journal} {Journal of Statistical Mechanics: Theory and
  Experiment}\ }\textbf {\bibinfo {volume} {2010}},\ \bibinfo {pages} {P02014}
  (\bibinfo {year} {2010})}\BibitemShut {NoStop}%
\bibitem [{\citenamefont {Mayo}\ \emph {et~al.}(2021)\citenamefont {Mayo},
  \citenamefont {Fan}, \citenamefont {Chern},\ and\ \citenamefont {del
  Campo}}]{Adolfo2021}%
  \BibitemOpen
  \bibfield  {author} {\bibinfo {author} {\bibfnamefont {J.~J.}\ \bibnamefont
  {Mayo}}, \bibinfo {author} {\bibfnamefont {Z.}~\bibnamefont {Fan}}, \bibinfo
  {author} {\bibfnamefont {G.-W.}\ \bibnamefont {Chern}},\ and\ \bibinfo
  {author} {\bibfnamefont {A.}~\bibnamefont {del Campo}},\ }\bibfield  {title}
  {\bibinfo {title} {Distribution of kinks in an ising ferromagnet after
  annealing and the generalized {K}ibble-{Z}urek mechanism},\ }\href
  {https://doi.org/10.1103/PhysRevResearch.3.033150} {\bibfield  {journal}
  {\bibinfo  {journal} {Phys. Rev. Research}\ }\textbf {\bibinfo {volume}
  {3}},\ \bibinfo {pages} {033150} (\bibinfo {year} {2021})}\BibitemShut
  {NoStop}%
\bibitem [{\citenamefont {Wang}\ \emph {et~al.}(2006)\citenamefont {Wang},
  \citenamefont {Nisoli}, \citenamefont {Freitas}, \citenamefont {Li},
  \citenamefont {McConville}, \citenamefont {Cooley}, \citenamefont {Lund},
  \citenamefont {Samarth}, \citenamefont {Leighton}, \citenamefont {Crespi},\
  and\ \citenamefont {Schiffer}}]{Wang06}%
  \BibitemOpen
  \bibfield  {author} {\bibinfo {author} {\bibfnamefont {R.~F.}\ \bibnamefont
  {Wang}}, \bibinfo {author} {\bibfnamefont {C.}~\bibnamefont {Nisoli}},
  \bibinfo {author} {\bibfnamefont {R.~S.}\ \bibnamefont {Freitas}}, \bibinfo
  {author} {\bibfnamefont {J.}~\bibnamefont {Li}}, \bibinfo {author}
  {\bibfnamefont {W.}~\bibnamefont {McConville}}, \bibinfo {author}
  {\bibfnamefont {B.~J.}\ \bibnamefont {Cooley}}, \bibinfo {author}
  {\bibfnamefont {M.~S.}\ \bibnamefont {Lund}}, \bibinfo {author}
  {\bibfnamefont {N.}~\bibnamefont {Samarth}}, \bibinfo {author} {\bibfnamefont
  {C.}~\bibnamefont {Leighton}}, \bibinfo {author} {\bibfnamefont {V.~H.}\
  \bibnamefont {Crespi}},\ and\ \bibinfo {author} {\bibfnamefont
  {P.}~\bibnamefont {Schiffer}},\ }\bibfield  {title} {\bibinfo {title}
  {Artificial `spin ice' in a geometrically frustrated lattice of nanoscale
  ferromagnetic islands},\ }\href {https://doi.org/10.1038/nature04447}
  {\bibfield  {journal} {\bibinfo  {journal} {Nature}\ }\textbf {\bibinfo
  {volume} {439}},\ \bibinfo {pages} {303} (\bibinfo {year}
  {2006})}\BibitemShut {NoStop}%
\bibitem [{\citenamefont {Nisoli}\ \emph {et~al.}(2013)\citenamefont {Nisoli},
  \citenamefont {Moessner},\ and\ \citenamefont {Schiffer}}]{Nisoli13}%
  \BibitemOpen
  \bibfield  {author} {\bibinfo {author} {\bibfnamefont {C.}~\bibnamefont
  {Nisoli}}, \bibinfo {author} {\bibfnamefont {R.}~\bibnamefont {Moessner}},\
  and\ \bibinfo {author} {\bibfnamefont {P.}~\bibnamefont {Schiffer}},\
  }\bibfield  {title} {\bibinfo {title} {Colloquium: Artificial spin ice:
  Designing and imaging magnetic frustration},\ }\href
  {https://doi.org/10.1103/RevModPhys.85.1473} {\bibfield  {journal} {\bibinfo
  {journal} {Rev. Mod. Phys.}\ }\textbf {\bibinfo {volume} {85}},\ \bibinfo
  {pages} {1473} (\bibinfo {year} {2013})}\BibitemShut {NoStop}%
\bibitem [{\citenamefont {Chern}\ \emph {et~al.}(2014)\citenamefont {Chern},
  \citenamefont {Reichhardt},\ and\ \citenamefont {Nisoli}}]{Chern14}%
  \BibitemOpen
  \bibfield  {author} {\bibinfo {author} {\bibfnamefont {G.-W.}\ \bibnamefont
  {Chern}}, \bibinfo {author} {\bibfnamefont {C.}~\bibnamefont {Reichhardt}},\
  and\ \bibinfo {author} {\bibfnamefont {C.}~\bibnamefont {Nisoli}},\
  }\bibfield  {title} {\bibinfo {title} {{Realizing three-dimensional
  artificial spin ice by stacking planar nano-arrays}},\ }\href
  {https://doi.org/10.1063/1.4861118} {\bibfield  {journal} {\bibinfo
  {journal} {Applied Physics Letters}\ }\textbf {\bibinfo {volume} {104}},\
  \bibinfo {pages} {013101} (\bibinfo {year} {2014})}\BibitemShut {NoStop}%
\bibitem [{\citenamefont {Perrin}\ \emph {et~al.}(2016)\citenamefont {Perrin},
  \citenamefont {Canals},\ and\ \citenamefont {Rougemaille}}]{Perrin16}%
  \BibitemOpen
  \bibfield  {author} {\bibinfo {author} {\bibfnamefont {Y.}~\bibnamefont
  {Perrin}}, \bibinfo {author} {\bibfnamefont {B.}~\bibnamefont {Canals}},\
  and\ \bibinfo {author} {\bibfnamefont {N.}~\bibnamefont {Rougemaille}},\
  }\bibfield  {title} {\bibinfo {title} {Extensive degeneracy, coulomb phase
  and magnetic monopoles in artificial square ice},\ }\href
  {https://doi.org/10.1038/nature20155} {\bibfield  {journal} {\bibinfo
  {journal} {Nature}\ }\textbf {\bibinfo {volume} {540}},\ \bibinfo {pages}
  {410} (\bibinfo {year} {2016})}\BibitemShut {NoStop}%
\bibitem [{\citenamefont {Lib\'al}\ \emph {et~al.}(2006)\citenamefont
  {Lib\'al}, \citenamefont {Reichhardt},\ and\ \citenamefont
  {Reichhardt}}]{Libal06}%
  \BibitemOpen
  \bibfield  {author} {\bibinfo {author} {\bibfnamefont {A.}~\bibnamefont
  {Lib\'al}}, \bibinfo {author} {\bibfnamefont {C.}~\bibnamefont
  {Reichhardt}},\ and\ \bibinfo {author} {\bibfnamefont {C.~J.~O.}\
  \bibnamefont {Reichhardt}},\ }\bibfield  {title} {\bibinfo {title} {Realizing
  colloidal artificial ice on arrays of optical traps},\ }\href
  {https://doi.org/10.1103/PhysRevLett.97.228302} {\bibfield  {journal}
  {\bibinfo  {journal} {Phys. Rev. Lett.}\ }\textbf {\bibinfo {volume} {97}},\
  \bibinfo {pages} {228302} (\bibinfo {year} {2006})}\BibitemShut {NoStop}%
\bibitem [{\citenamefont {Ortiz-Ambriz}\ and\ \citenamefont
  {Tierno}(2016)}]{Ortiz16}%
  \BibitemOpen
  \bibfield  {author} {\bibinfo {author} {\bibfnamefont {A.}~\bibnamefont
  {Ortiz-Ambriz}}\ and\ \bibinfo {author} {\bibfnamefont {P.}~\bibnamefont
  {Tierno}},\ }\bibfield  {title} {\bibinfo {title} {Engineering of frustration
  in colloidal artificial ices realized on microfeatured grooved lattices},\
  }\href {https://doi.org/10.1038/ncomms10575} {\bibfield  {journal} {\bibinfo
  {journal} {Nature Communications}\ }\textbf {\bibinfo {volume} {7}},\
  \bibinfo {pages} {10575} (\bibinfo {year} {2016})}\BibitemShut {NoStop}%
\bibitem [{\citenamefont {Ortiz-Ambriz}\ \emph {et~al.}(2019)\citenamefont
  {Ortiz-Ambriz}, \citenamefont {Nisoli}, \citenamefont {Reichhardt},
  \citenamefont {Reichhardt},\ and\ \citenamefont {Tierno}}]{Ortiz19}%
  \BibitemOpen
  \bibfield  {author} {\bibinfo {author} {\bibfnamefont {A.}~\bibnamefont
  {Ortiz-Ambriz}}, \bibinfo {author} {\bibfnamefont {C.}~\bibnamefont
  {Nisoli}}, \bibinfo {author} {\bibfnamefont {C.}~\bibnamefont {Reichhardt}},
  \bibinfo {author} {\bibfnamefont {C.~J.~O.}\ \bibnamefont {Reichhardt}},\
  and\ \bibinfo {author} {\bibfnamefont {P.}~\bibnamefont {Tierno}},\
  }\bibfield  {title} {\bibinfo {title} {Colloquium: Ice rule and emergent
  frustration in particle ice and beyond},\ }\href
  {https://doi.org/10.1103/RevModPhys.91.041003} {\bibfield  {journal}
  {\bibinfo  {journal} {Rev. Mod. Phys.}\ }\textbf {\bibinfo {volume} {91}},\
  \bibinfo {pages} {041003} (\bibinfo {year} {2019})}\BibitemShut {NoStop}%
\bibitem [{\citenamefont {Francuz}\ \emph {et~al.}(2016)\citenamefont
  {Francuz}, \citenamefont {Dziarmaga}, \citenamefont {Gardas},\ and\
  \citenamefont {Zurek}}]{Francuz16}%
  \BibitemOpen
  \bibfield  {author} {\bibinfo {author} {\bibfnamefont {A.}~\bibnamefont
  {Francuz}}, \bibinfo {author} {\bibfnamefont {J.}~\bibnamefont {Dziarmaga}},
  \bibinfo {author} {\bibfnamefont {B.}~\bibnamefont {Gardas}},\ and\ \bibinfo
  {author} {\bibfnamefont {W.~H.}\ \bibnamefont {Zurek}},\ }\bibfield  {title}
  {\bibinfo {title} {Space and time renormalization in phase transition
  dynamics},\ }\href {https://doi.org/10.1103/PhysRevB.93.075134} {\bibfield
  {journal} {\bibinfo  {journal} {Phys. Rev. B}\ }\textbf {\bibinfo {volume}
  {93}},\ \bibinfo {pages} {075134} (\bibinfo {year} {2016})}\BibitemShut
  {NoStop}%
\bibitem [{\citenamefont {Jeong}\ \emph {et~al.}(2019)\citenamefont {Jeong},
  \citenamefont {Kim},\ and\ \citenamefont {Lee}}]{Lee2019}%
  \BibitemOpen
  \bibfield  {author} {\bibinfo {author} {\bibfnamefont {K.}~\bibnamefont
  {Jeong}}, \bibinfo {author} {\bibfnamefont {B.}~\bibnamefont {Kim}},\ and\
  \bibinfo {author} {\bibfnamefont {S.~J.}\ \bibnamefont {Lee}},\ }\bibfield
  {title} {\bibinfo {title} {Growth kinetics of the two-dimensional ising model
  with finite cooling rates},\ }\href
  {https://doi.org/10.1103/PhysRevE.99.022113} {\bibfield  {journal} {\bibinfo
  {journal} {Phys. Rev. E}\ }\textbf {\bibinfo {volume} {99}},\ \bibinfo
  {pages} {022113} (\bibinfo {year} {2019})}\BibitemShut {NoStop}%
\bibitem [{\citenamefont {Wills}\ \emph {et~al.}(2002)\citenamefont {Wills},
  \citenamefont {Ballou},\ and\ \citenamefont {Lacroix}}]{Wills02}%
  \BibitemOpen
  \bibfield  {author} {\bibinfo {author} {\bibfnamefont {A.~S.}\ \bibnamefont
  {Wills}}, \bibinfo {author} {\bibfnamefont {R.}~\bibnamefont {Ballou}},\ and\
  \bibinfo {author} {\bibfnamefont {C.}~\bibnamefont {Lacroix}},\ }\bibfield
  {title} {\bibinfo {title} {Model of localized highly frustrated
  ferromagnetism: The kagom\'e spin ice},\ }\href
  {https://doi.org/10.1103/PhysRevB.66.144407} {\bibfield  {journal} {\bibinfo
  {journal} {Phys. Rev. B}\ }\textbf {\bibinfo {volume} {66}},\ \bibinfo
  {pages} {144407} (\bibinfo {year} {2002})}\BibitemShut {NoStop}%
\bibitem [{\citenamefont {M\"oller}\ and\ \citenamefont
  {Moessner}(2009)}]{Moller09}%
  \BibitemOpen
  \bibfield  {author} {\bibinfo {author} {\bibfnamefont {G.}~\bibnamefont
  {M\"oller}}\ and\ \bibinfo {author} {\bibfnamefont {R.}~\bibnamefont
  {Moessner}},\ }\bibfield  {title} {\bibinfo {title} {Magnetic multipole
  analysis of kagome and artificial spin-ice dipolar arrays},\ }\href
  {https://doi.org/10.1103/PhysRevB.80.140409} {\bibfield  {journal} {\bibinfo
  {journal} {Phys. Rev. B}\ }\textbf {\bibinfo {volume} {80}},\ \bibinfo
  {pages} {140409} (\bibinfo {year} {2009})}\BibitemShut {NoStop}%
\bibitem [{\citenamefont {Chern}\ \emph {et~al.}(2011)\citenamefont {Chern},
  \citenamefont {Mellado},\ and\ \citenamefont {Tchernyshyov}}]{Chern11}%
  \BibitemOpen
  \bibfield  {author} {\bibinfo {author} {\bibfnamefont {G.-W.}\ \bibnamefont
  {Chern}}, \bibinfo {author} {\bibfnamefont {P.}~\bibnamefont {Mellado}},\
  and\ \bibinfo {author} {\bibfnamefont {O.}~\bibnamefont {Tchernyshyov}},\
  }\bibfield  {title} {\bibinfo {title} {Two-stage ordering of spins in dipolar
  spin ice on the kagome lattice},\ }\href
  {https://doi.org/10.1103/PhysRevLett.106.207202} {\bibfield  {journal}
  {\bibinfo  {journal} {Phys. Rev. Lett.}\ }\textbf {\bibinfo {volume} {106}},\
  \bibinfo {pages} {207202} (\bibinfo {year} {2011})}\BibitemShut {NoStop}%
\bibitem [{\citenamefont {Fan}\ and\ \citenamefont {Chern}(2023)}]{Fan23}%
  \BibitemOpen
  \bibfield  {author} {\bibinfo {author} {\bibfnamefont {Z.}~\bibnamefont
  {Fan}}\ and\ \bibinfo {author} {\bibfnamefont {G.-W.}\ \bibnamefont
  {Chern}},\ }\href@noop {} {\bibinfo {title} {Nonequilibrium generation of
  charge defects in kagome spin ice under slow cooling}} (\bibinfo {year}
  {2023}),\ \Eprint {https://arxiv.org/abs/2306.17036} {arXiv:2306.17036
  [cond-mat.stat-mech]} \BibitemShut {NoStop}%
\bibitem [{\citenamefont {Hamp}\ \emph {et~al.}(2015)\citenamefont {Hamp},
  \citenamefont {Chandran}, \citenamefont {Moessner},\ and\ \citenamefont
  {Castelnovo}}]{Hamp15}%
  \BibitemOpen
  \bibfield  {author} {\bibinfo {author} {\bibfnamefont {J.}~\bibnamefont
  {Hamp}}, \bibinfo {author} {\bibfnamefont {A.}~\bibnamefont {Chandran}},
  \bibinfo {author} {\bibfnamefont {R.}~\bibnamefont {Moessner}},\ and\
  \bibinfo {author} {\bibfnamefont {C.}~\bibnamefont {Castelnovo}},\ }\bibfield
   {title} {\bibinfo {title} {Emergent coulombic criticality and
  {K}ibble-{Z}urek scaling in a topological magnet},\ }\href
  {https://doi.org/10.1103/PhysRevB.92.075142} {\bibfield  {journal} {\bibinfo
  {journal} {Phys. Rev. B}\ }\textbf {\bibinfo {volume} {92}},\ \bibinfo
  {pages} {075142} (\bibinfo {year} {2015})}\BibitemShut {NoStop}%
\end{thebibliography}%

\end{document}